\documentclass[prb,twocolumn,showpacs,superscriptaddress,amsmath,amssymb]{revtex4-1}

\usepackage{graphicx}% Include figure files
\usepackage{dcolumn}% Align table columns on decimal point
\usepackage{bm}% bold math      
\usepackage{xcolor}

\begin{document}

\title{Evolution of short-range magnetic correlations in ferromagnetic Ni-V alloys} 
                                                                                                                                                         
\author{Shiva Bhattarai}
\affiliation{Physics Department, Kent State University, Kent OH 44242, USA}
\author{Hind Adawi}
\affiliation{Physics Department, Kent State University, Kent OH 44242, USA}
\affiliation{Department of Physics, Jazan University, Jazan 45142, Kingdom of Saudi-Arabia}
\author{Jean-Guy Lussier}
\affiliation{Physics Department, Kent State University, Kent OH 44242, USA}
\author{Adane Gebretsadik}
\email{Present address: Intel Chandler AZ, USA} 
\affiliation{Physics Department, Kent State University, Kent OH 44242, USA}
\author{Maxim Dzero}
\affiliation{Physics Department, Kent State University, Kent OH 44242, USA}
\author{Kathryn L. Krycka}
\affiliation{NIST Center of Neutron Research, National Institute of Standards and Technology, Gaithersburg MD 20899, USA}
\author{Almut Schroeder}%
\email{aschroe2@kent.edu}
\affiliation{Physics Department, Kent State University, Kent OH 44242, USA}

%\date{\today}

\begin{abstract}
We experimentally study how the magnetic correlations develop in a binary alloy close to the ferromagnetic quantum critical point with small-angle neutron scattering (SANS). Upon alloying the itinerant ferromagnet nickel with vanadium, the ferromagnetic order is continuously suppressed. The critical temperature $T_c$ vanishes when vanadium concentrations reach the critical value of $x_c=0.116$ indicating a quantum critical point separating the ferromagnetic and paramagnetic phases. Earlier magnetization and $\mu$SR data have indicated the presence of magnetic inhomogeneities in Ni$_{1-x}$V$_x$ and, in particular, recognize the magnetic clusters close to $x_c$, on the paramagnetic and on the ferromagnetic sides with nontrivial dynamical properties [R. Wang {\it et al.}, Phys. Rev. Lett. {\bf 118}, 267202 (2017)]. 
We present the results of SANS study with full polarization analysis of polycrystalline Ni$_{1-x}$V$_x$ samples with $x=0.10$ and $x=0.11$  with low critical temperatures $T_c<50$ K. For both Ni-V samples close to $x_c$ we find isotropic magnetic short-range correlations on the nanometer-scale persisting at low temperatures. They are suppressed gradually in higher magnetic fields. In addition, signatures of long-range ordered magnetic domains are present below $T_c$. The fraction of these magnetic clusters embedded in the ferromagnetic ordered phase grows toward $x_c$ and agrees well with the cluster fraction estimate from the magnetization and $\mu$SR data. Our SANS studies provide new insights into the nature of the inhomogeneities in a ferromagnetic alloy close to a quantum critical point.

\end{abstract}

\maketitle

%%%%%%%%%%%%%%%%%%%%%%%%%%%%%%%%%%%%%%%%%%%%%%%%%%%%%%%%%%%%%%%%%%%%%%%%%%%%%%%%%%%%%%%%%%%%%%%%%%%%%
% Introduction 
%%%%%%%%%%%%%%%%%%%%%%%%%%%%%%%%%%%%%%%%%%%%%%%%%%%%%%%%%%%%%%%%%%%%%%%%%%%%%%%%%%%%%%%%%%%%%%%%%%%%%
\section{\label{sec:intro}Introduction}
Ferromagnetic order emerging as a result of electron-electron interactions in metals has been studied extensively during the last several decades. 
In clean metals if the spin-exchange interactions between the itinerant electrons are strong enough, then the ferromagnetic order develops at some finite temperature.\cite{stoner1938,Stoner1947,Pomeranchuk} For example, for Ni the critical temperature of the ferromagnetic transition is $T_c\approx 630$ K. Introduction of disorder by alloying is a way to suppress the ferromagnetic transition, so that the critical temperature becomes zero at some finite concentration of impurity atoms $x_c$. For example, in Ni$_{1-x}$V$_x$ alloys the critical concentration is $x_c\approx 0.116$.\cite{Ubaid2010} 

When $x=x_c$ a quantum phase transition (QPT) between the ferromagnetic and paramagnetic states is expected to occur. Since the symmetry of the ground state changes at the quantum critical point (QCP), the evolving quantum critical fluctuations affect the vicinity of the QCP, including finite temperatures that lead to unusual thermal properties.\cite{Sachdev2011book} Since the transition is driven by spin-exchange interactions, which for each realization of disorder become nonlocal and random, QPTs are expected to be affected by disorder more likely than thermal phase transitions.\cite{Vojta2014} 

%One important consequence of this fact is that 
Under the proper conditions random quenched disorder in a metallic system can produce a quantum Griffiths phase \cite{Vojta2006,Vojta2013} that is usually recognized by  anomalous thermodynamic properties \cite{Vojta2013} close to the QCP. In this case, a distribution of magnetic fluctuations with different time scales including very slow rare regions \cite{Vojta2006} dominates the magnetic response. Specifically, experimental signatures of the Griffiths phase are power laws with non-universal exponents\cite{Vojta10} in thermodynamic responses. These power laws are observed in many systems \cite{Stewart2001,Ubaid2010}, but the responsible fluctuations with different time and and length scales have not been demonstrated directly. 
In general, disordered alloys close to a ferromagnetic QCP are recognized by unusual scaling behavior \cite{Huang2016,Sales2017,Mishra2020}.  Specific effective exponents are predicted for disordered FMs. \cite{KB2015,Brando2016}
Note that QCPs in FMs are rare; more often new phases emerge when $T_c$ gets suppressed.\cite{Brando2016} Typically, clean FMs rather present a first-order transition\cite{Brando2016} without critical fluctuations.
So remarkably, introducing disorder is a unique route for a QCP in itinerant FM such as Ni. The characterization of the critical fluctuation spectrum should reveal the nature of this special point.

Lastly, we note that the conclusion that only disorder allows to the QCP in itinerant ferromagnets is a simplification as 
 alternative mechanisms have been recently proposed to explain a ferromagnetic QCP\cite{KB2020} with critical fluctuations. For example, the experimental signatures of the quantum critical fluctuations were reported in chemically tuned Ni-Rh\cite{Huang2020} and pressure-tuned CeRh$_6$Ge$_4$\cite{Shen2020,Shu2021} from ferromagnetic systems where disorder is considered negligible.

As we have mentioned above, Ni$_{1-x}$V$_x$ is an example of a disordered metallic system which exhibits the signatures of a ferromagnetic QCP indicating the presence of quantum critical fluctuations. In this paper, we present the results of the measurements which directly probe quantum critical magnetic fluctuations in the vicinity of the ferromagnetic QPT driven by disorder. In particular, we use
small-angle neutron scattering (SANS) to probe the magnetic microstructure and magnetic inhomogeneity on the mesoscopic length scale of interest from a few to a hundred nanometers. \cite{mSANSrev2019} We are motivated by the fact that SANS successfully revealed the correlation length change of critical fluctuations close to a phase transition in several magnetic alloys. \cite{Collins1989,Murani1976,Verbeek1980,Burke1983,Grigoriev2001}
We present the experimental evidence of magnetic correlations at various length scales in a ferromagnetic Ni alloy introduced by random atomic substitution of Ni by V.  These findings are compatible with a quantum Griffiths phase.  
\begin{figure}
\includegraphics[width=\columnwidth]{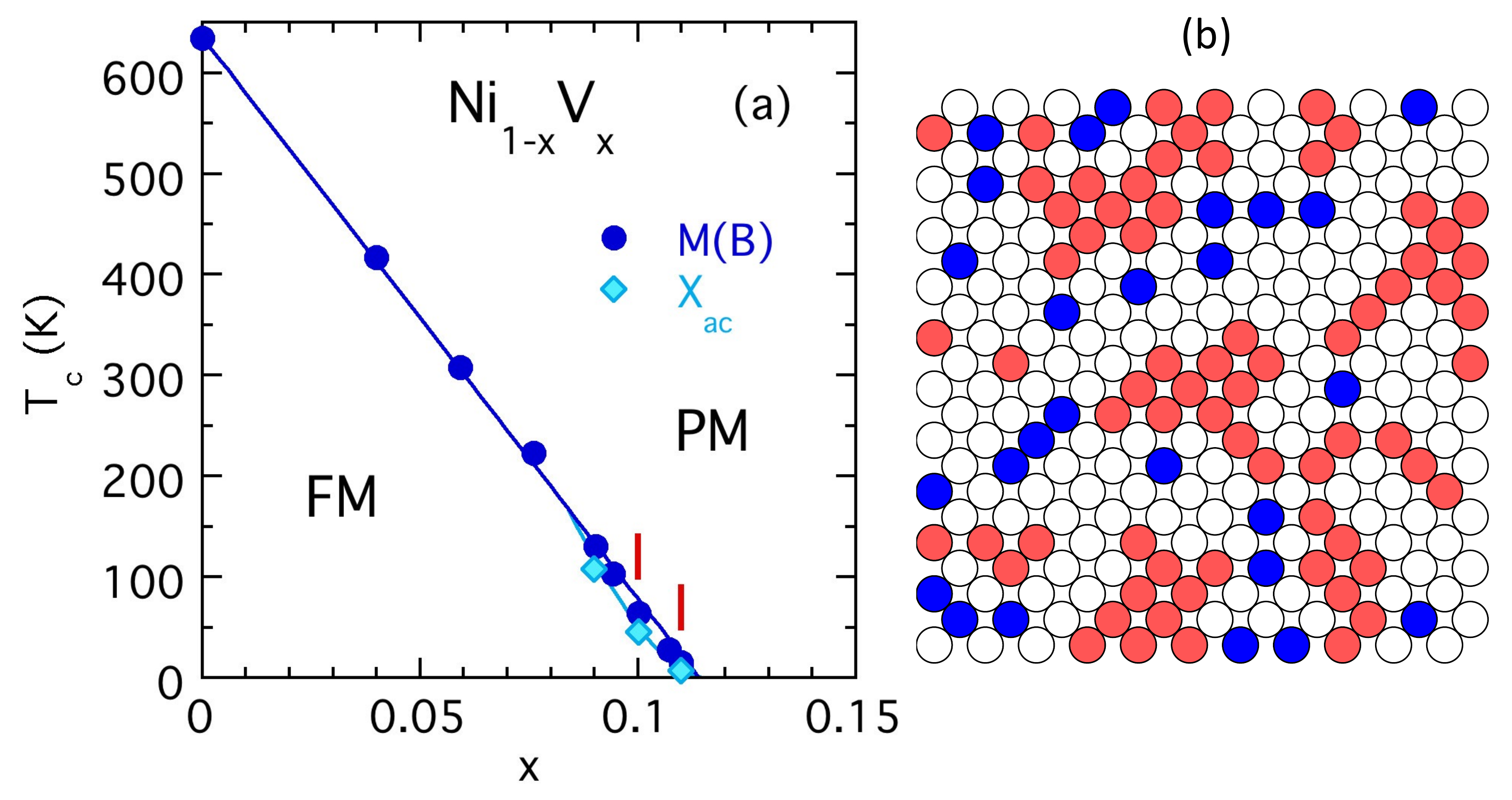}% phase diagram
\caption{(a) Magnetic phase diagram of Ni$_{1-x}$V$_x$ displaying the critical temperature $T_c$  vs. vanadium concentration $x$ separating the ferromagnetic ordered phase (FM) from the paramagnetic phase (PM) leading to a quantum critical point.\cite{Wang2017} The red markers indicate the samples for this investigation. The right panel (b) demonstrates the magnetic inhomogeneities in Ni$_{1-x}$V$_x$ with $x=0.10$ in an fcc lattice plane: it shows how randomly placed V atoms (in blue) separate Ni-rich regions (in red) responsible for magnetism. Since V affects more the moment of the local Ni neighbors, we distinguish these close Ni-sites (in white) from the rest, the more distant Ni atoms (in red). 
} 
\label{fig:phase}
\end{figure}
%

%
% %%%%%%%%%%%%%%%%%%%%%%%%%%%%%%%%%
% Experimentals
%%%%%%%%%%%%%%%%%%%%%%%%%%%%%%%%%%%
%
\section{\label{sec:methods}Experimental Methods}
\subsection{\label{sec:NiVintro} Ni-V}
%As we have mentioned above, we will be studying Ni$_{1-x}$V$_x$ alloys. 
The alloys of Ni$_{1-x}$V$_x$ provide a good platform to study a random disordered QPT not only because of their low magnetic anisotropy and the ``large defects" caused by vanadium, but also because they form indeed good solid solutions with random atomic disorder. 
In addition, when the vanadium concentration reaches the value of $x_c$ = $0.116$ these binary alloys exhibit a quantum phase transition (QPT) from the ferromagnetic (FM) to a paramagnetic (PM) phase with a quantum critical point (QCP)\cite{Ubaid2010}. Fig.\ \ref{fig:phase}(a) shows the phase diagram. It is known that V affects the electronic state of the Ni neighbors and causes therefore large magnetic inhomogeneities \cite{Collins1965} and an effective reduction of the magnetic moment with low $x_c$\cite{Friedel1958}. In Fig.\ \ref{fig:phase}(b) we present a schematic view of a lattice plane with V on random lattice sites. Assuming that V causes a moment reduction of neighboring Ni, the undisturbed magnetic Ni network becomes inhomogeneous. 

Previous studies reveal some signatures of a quantum Griffiths phase. Specifically, for $x\approx x_c$ the magnetization data $M(B)$ show the non-universal power law dependencies.\cite{Ubaid2010,Wang2017} This is further supported by  $\mu$SR data recognizing a field distribution in the samples\cite{Schroeder2014}, and the ferromagnetic alloys with values of $x$ close to $x_c$ include a dynamic contribution to $M(B)$ besides the static contribution.\cite{Wang2017} 

 All samples keep the simple fcc crystal structure. The polycrystalline samples were arc-melted and annealed at high temperatures and cooled down fast to maintain the random atomic placement of the V on the fcc lattice sites at low temperatures. A structural study (PDF analysis from wide-angle neutron diffraction data \cite{Wang2017,PDF}) confirmed the random atomic distribution of V and the otherwise unchanged crystalline fcc lattice at low temperatures.  
 
\subsection{\label{sec:exp}Experimental Details}
\begin{figure}
\includegraphics[width=\columnwidth]{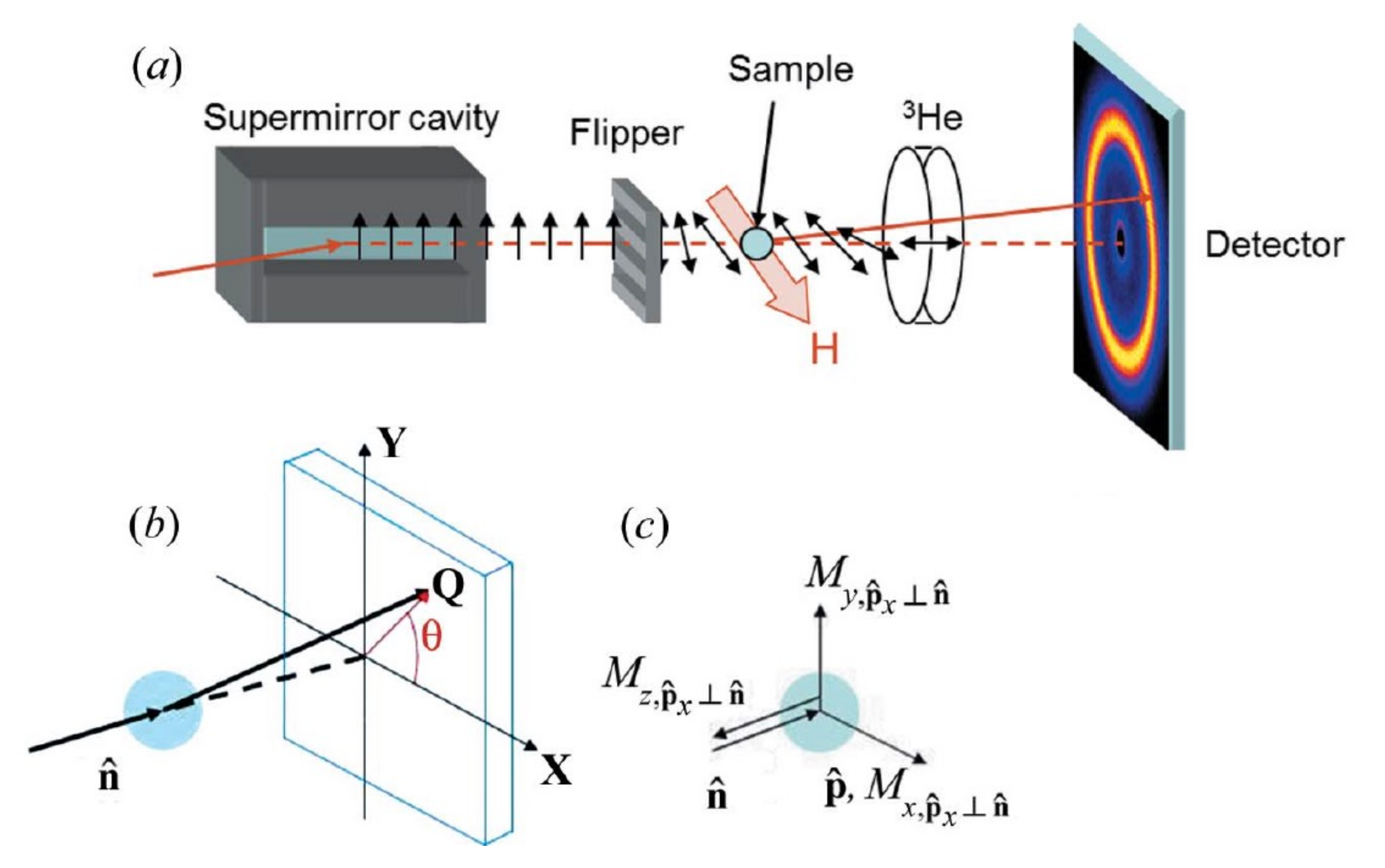}% Fig2 sans
\caption{(a) Polarized SANS set up with polarizing supermirror, spin flipper, sample in a cryostat with a magnetic field perpendicular to the neutron beam, a $^3$He analyzer, and a position sensitive detector. Arrows indicate neutron polarization direction. (b) Coordinate axes with beam direction $\hat{n}$ along $z$ defining angle $\theta$ in the $xy$ plane of the detector. (c) Present set up of field and polarization direction $\hat{p}$ along $x$ (from Ref. \onlinecite{Krycka2012})} 
\label{fig:sans}
\end{figure}

We use the same polycrystalline samples prepared for optimized random atomic distribution as studied before by different methods.\cite{Wang2017,PDF,Ubaid2010} For the small-angle neutron scattering (SANS) study we chose the concentrations $x=0.10$, $x=0.11$ and $x=0$. The samples with $x=0.11$ are made with $^{58}$Ni, while the others contain natural Ni isotope mixtures, which yield different nuclear neutron cross sections of Ni. 
%(e.g. for thermal neutrons $sigma_{coh} (Ni) = barn$ $\sigma_{coh} Ni^{58}= barn$) 
The SANS experiments were performed at the instruments NG7SANS\cite{Glinka1998} and VSANS\cite{VSANS}, at the NIST Center for Neutron Research (NCNR) at the National Institute of Standards and Technology and at the instrument GPSANS, at the high flux isotope reactor (HFIR) at Oak Ridge National Laboratory. We focus here most on the SANS experiments that allow a full polarization analysis from NG7SANS. Several 3 mm diameter pellets of each concentration were wrapped in Al foil and placed on a Cd-mask framed Al-sample holder attached to the cold plate of the cryostat. To cover a wave vector range of $Q=$ (0.06 - 1) nm$^{-1}$ with neutron wavelengths of 0.55 nm and 0.75 nm, the SANS intensity was collected in the $xy$ plane on a 2D detector at different sample to detector distances (from 2m to 11m). We obtained the different polarized cross sections, e.g. the non-spin flip (NSF) scattering with unchanged polarization state of the neutrons (DD and UU) and spin flip (SF) scattering with reverse polarization state (DU and UD) from the sample, using the super mirror polarizer and $^3$He-cell as a spin analyzer as stated in detail in Refs. \onlinecite{Krycka2009,Chen2009}.   
See set up in Fig. \ref{fig:sans}. U, D refer to the up, down aligned neutron spins with regard to the axis of neutron polarization determined by the external magnetic field. The magnetic field was applied in the $x$ direction ($B_{min}$ = 7 mT, $B_{max}$ = 1.5 T) perpendicular to the beam ($\parallel$z). $\theta$ indicates the azimuthal angle within the $xy$ plane, with $\theta=0^\circ$ in the horizontal $x$-direction and $\theta=90^\circ$ in vertical $y$ direction of the detector. Most of the data were reduced and analyzed with the IGOR software\cite{Kline2006}.

\subsection{\label{sec:pol}Polarized SANS Analysis}
The full polarized SANS (PASANS) technique traces the neutron spin before it enters and after it leaves the sample separately.  The four different cross sections (DD, UU, DU, UD) offer the advantage of separating the small magnetic scattering from overwhelming nuclear scattering in our samples.\cite{Moon1969} Here we use in particular the angle dependence \cite{Krycka2012} of the signal in the $xy$ plane of the detector to resolve magnetic components $M_x^2$, $M_y^2$, $M_x^2$, with a magnetic field $B_x$ perpendicular to the beam. We apply two methods to best trace the magnetic signal. First we collect the pure SF signal (DU+UD) that ideally selects ``magnetic" scattering. But it contains other ``background" (BG) contributions due to incoherent scattering from the sample and other sources. A small SF contribution of V nuclear spins is also included that does not show any angle dependence.\cite{Moon1969, Schroeder2020}  

The NSF signal (DD+UU) is only used in limited cases to extract some strong longitudinal component $M_x^2$ when the nuclear contribution $N^2$ and its sample variation are not dominating. Otherwise, we use the flipper contrast of NSF data called DIF of fully polarized (DD-UU) or half polarized data (D-U) to recognize the anisotropic $M_x$ component of the mixed term $2NM_x$ that contains a strong nuclear contribution. For our geometry we do not expect and do not see any indication of anisotropy in the transverse direction of the field, so that we keep $M_y=M_z=0$, while $M_x>0$.
 
Note that the intensity shown is not calibrated, it is consistent for each sample. 
Different number and size of pellets and the different cross sections of  $^{58}$Ni and of natural Ni lead to different responses for both samples with $x=0.11$ and $x=0.10$, respectively.
We omit also any proportionality factor in the equation (more details and general forms are given in Ref. \onlinecite{Krycka2012}). 

The angular dependence for the SF response is therefore
\begin{equation}
SF(\theta)= M_z^2 + M_y^2\cos^4\theta+ M_x^2\cos^2\theta \sin^2\theta + BG_{SF}.
\label{equation:SFA} 
\end{equation}
For the case of isotropic magnetic correlations where $M_x^2=M_y^2=M_z^2=\frac{1}{3} M_{tot}^2$, we expect a simple cosine-square response with amplitude of $\frac{1}{3} M_{tot}^2$:
\begin{equation}
SF(\theta)= M_y^2(1+ \cos^2\theta)+BG_{SF}.
\label{equation:SFA_cosine} 
\end{equation}
In particular, the SF contrast, the difference between SF data in a horizontal sector ($\theta=0^\circ$ and $180^\circ$) called SFH and a vertical sector ($\theta=\pm 90^\circ$) called SFV, each collected typically within $\delta \pm30^\circ$, produces the transverse component $M_y^2$ without any BG$_{SF}$. Ideally BG$_{SF}$ cancels out in the difference assuming the other contributions to SF($\theta$) are not angle dependent:
\begin{equation}
SF(\theta=0^\circ)-SF(\theta=90^\circ)=M_y^2(Q).
 \end{equation}
This SF contrast can be evaluated over the accessible $Q$-regime to study the magnetic correlation lengths.

In principle the total non-spin flip data NSF serve to extract the longitudinal component of the magnetic scattering, $M_x^2$, from the angular dependence. But the extra constant is not small and not angle independent due to the non homogeneous sample arrangement. The nuclear scattering of the sample $N^2$ dominates typically the response compared to other external BG and magnetic signals.
\begin{equation}
NSF_{\textrm{tot}}(\theta)= N^2+ M_x^2 \sin^4\theta+M_y^2 \cos^2\theta \sin^2\theta + BG_{NSF}.
\label{equation:NSFA} 
\end{equation}
For isotropic correlations we get a simple sine-square variation with amplitude $M_x^2=\frac{1}{3}M_{tot}^2$:
\begin{equation}
NSF(\theta)= N^2 + M_x^2\sin^2\theta + BG_{NSF}
\label{equation:NSFA_sine} 
\end{equation}

The different response between the two polarization directions (without registering spin flip), the NSF asymmetry or DIF (DD-UU or D-U) yields an interference term of nuclear and magnetic origin. It signals a weak contribution from a center with a net magnetic component along the $x$ direction $M_x$ in the presence of a strong nuclear contribution from the same center:
\begin{equation}
DIF(\theta)= 2NM_x \sin^2\theta   
\label{equation:DIF} 
\end{equation}
The vertical sector cut DIFV (collected at $\theta=\pm90^\circ$  within $\delta \pm 30^\circ$) gives the maximum signal $2NM_x $. We can trace this DIFV term over a large $Q$-regime resolving an anisotropic magnetic response $M_x$ from 0.06 nm$^{-1}$ to 1 nm$^{-1}$ in the Ni$_{1-x}$V$_x$ samples.

%%%%%%%%%%%%%%%%%%%%%%%%%%%%%%%%%%%%%%%%%%%%%%%%%%%%%%%%%%%%%%%%%%%%%%%%%%%%%%%%%%%%%%%%%%%%%%%%%%%%%
% Fluctuations
%%%%%%%%%%%%%%%%%%%%%%%%%%%%%%%%%%%%%%%%%%%%%%%%%%%%%%%%%%%%%%%%%%%%%%%%%%%%%%%%%%%%%%%%%%%%%%%%%%%%%
\section{\label{sec:results} Results of SANS Study}
We collected indication of disorder and clusters in Ni$_{1-x}$V$_x$ alloys \cite{Wang2017} related to a quantum Griffith phase. But any information of length and time scales of these unusual magnetic clusters 
%that are not expected in homogeneous ferromagnets,    
is still lacking. Initial SANS experiments demonstrated that some magnetic signal can be detected for samples with small vanadium concentration $x<x_c$, while the magnetic intensity in the paramagnetic regime ($x>x_c$) seems too small to be noticed with an averaged high-field magnetic moment of less than 0.01$\mu_B$/atom \cite{Ubaid2010}. More recent (unpolarized) SANS data collected at GPSANS show promising temperature and field dependent signals, and finally polarized SANS data can identify consistently a magnetic response.  Some data with preliminary analysis for the compound Ni$_{0.9}$V$_{0.10}$ with the higher critical temperature $T_c\approx50$ K are shown in Ref. \onlinecite{Schroeder2020}. Here we present the PASANS results of the samples closest to the critical point with the lowest $T_c$ in contrast to pure Ni. We show in detail the magnetic responses of Ni$_{0.89}$V$_{0.11}$ with $T_c$= 7 K and compare them with Ni$_{0.9}$V$_{0.10}$ with $T_c\approx50$ K to figure out the relevant magnetic correlations, their correlation length, and evolution toward $x_c$.

\subsection{\label{sec:SF10} Short-Range Correlations in Ni$_{0.90}$V$_{0.10}$}
The first challenge is to extract the small magnetic scattering from the large nuclear contribution of Ni. The background of ``non magnetic scattering" is high because of the high nuclear cross section of Ni compared to the small magnetic cross section due to the reduced average magnetic moment ($< 0.1\mu_B/{\textrm{Ni}}$) in these alloys. The additional grain boundary scattering observed in our polycrystalline samples dominates toward lower scattering vectors $Q$. We use the (azimuthal) angle dependence (in the $xy$ plane of the detector) of selected cross sections to resolve the magnetic components $M_x^2$, $M_y^2$, $M_z^2$ with a magnetic field $B_x$ perpendicular to the beam (along $z$). Section \ref{sec:pol} presents the relevant expressions for our set up shown in Fig.\,\ref{fig:sans}(b,c). 

\begin{figure}
\includegraphics[width=.9\columnwidth]{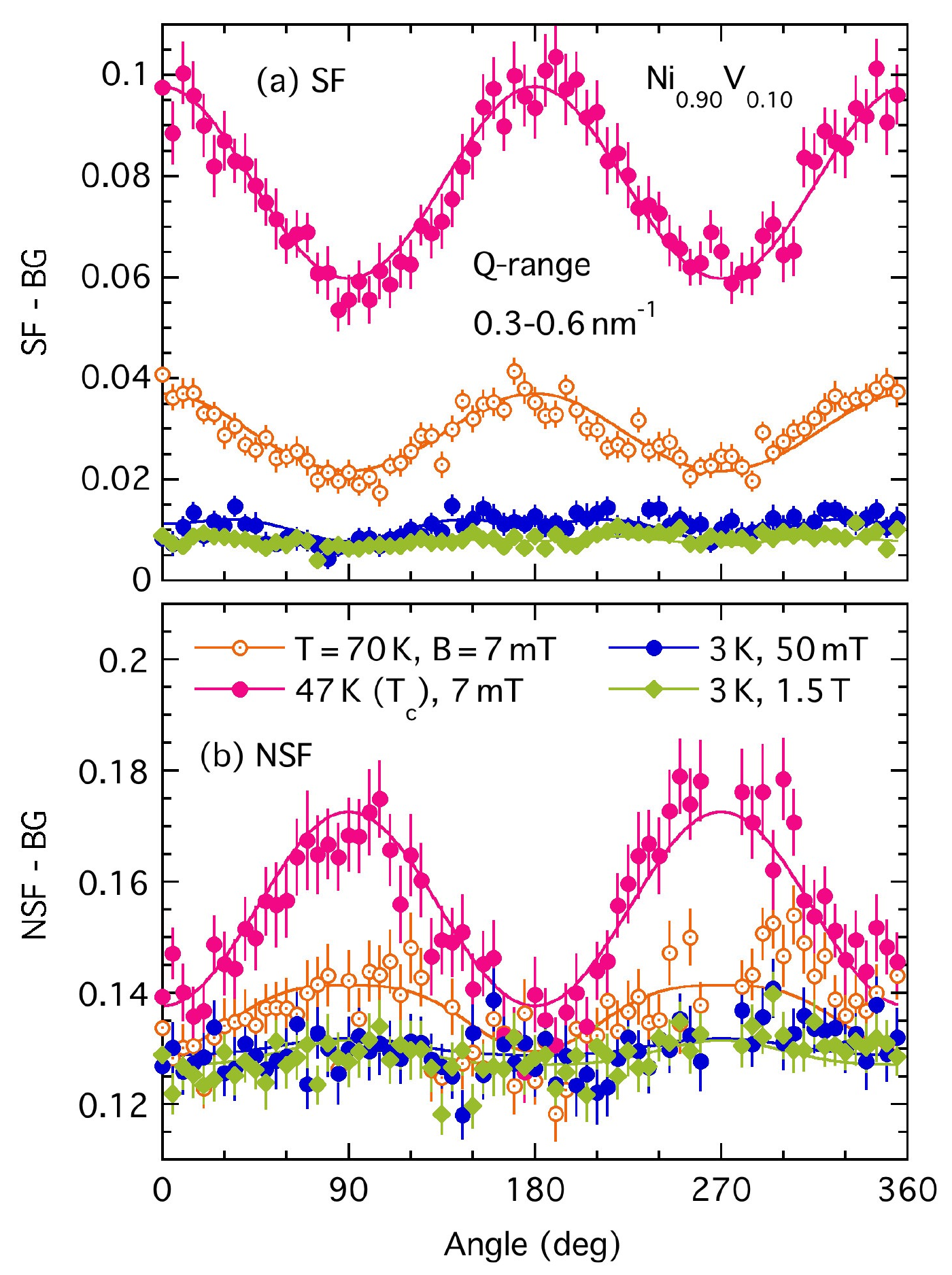}% SFA10 called before AnnularAverage10
\caption
{Neutron scattering intensity of Ni$_{0.90}$V$_{0.10}$ vs azimuthal angle $\theta$ collected in medium $Q$-range
of $(0.3-0.6)$ nm$^{-1}$ at different temperatures $T$ and magnetic fields $B$. The upper panel (a) presents the SF data, the lower panel (b) the NSF data with fit as solid line using
Eq.\ (\ref{equation:SFA}) in (a) and Eq.\ (\ref{equation:NSFA}) in (b). An angle independent background BG has been subtracted from the data, which varies with spin filter and $B$.}
\label{fig:SFA10}
\end{figure}

\begin{figure}
\includegraphics[width=\columnwidth]{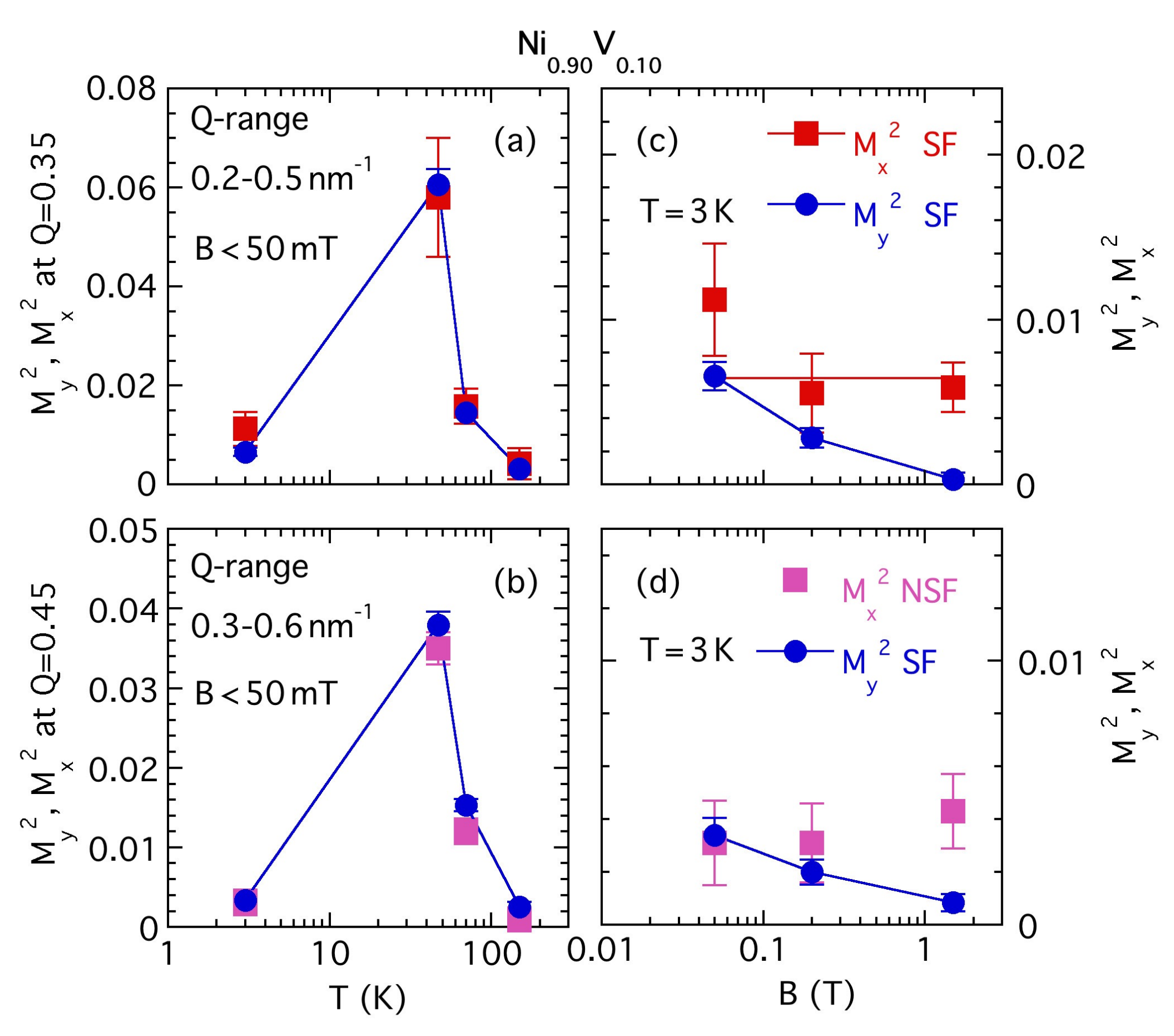}% para10
\caption{ 
Temperature $T$ and magnetic field $B$ dependence of magnetic transverse components $M_y^2$ and longitudinal component $M_x^2$ 
of Ni$_{0.90}$V$_{0.10}$. 
In the upper panel (a,c) $M_y^2$ and $M_x^2$ are extracted from fit  (Eq.\ (\ref{equation:SFA})) of SF($\theta$)  
at $Q$-range $(0.2-0.5)$ nm$^{-1}$.  
The lower panel (b,d) presents the main components from the data shown in Fig.\ \ref{fig:SFA10} at $Q$-range $(0.3-0.6)$ nm$^{-1}$: $M_y^2$ comes from fit of SF($\theta$) and $M_x^2$ from fit (Eq.\ (\ref{equation:NSFA})) of NSF($\theta$).
} 
\label{fig:x10MxMy}
\end{figure}

First, we focus on the spin flip (SF) signal to recognize the magnetic scattering. The angle dependence of SF($\theta$) as predicted by Eq.\ (\ref{equation:SFA}) further separates magnetic scattering from the background contribution and recognizes spin anisotropies. 
We distinguish a longitudinal magnetic component $M_x^2$ from transverse magnetic components $M_y^2$ and $M_z^2$ (with a wave vector $Q$ perpendicular to $z$). 
Fig.\ \ref{fig:SFA10} presents $SF(\theta)$ and the non-spin flip signal $NSF(\theta)$ of Ni$_{0.9}$V$_{0.10}$   in a medium $Q$-range ($0.45\pm0.15$) nm$^{-1}$. 
Both show significant variations in the angle $\theta$ most obvious for a temperature close to $T_c$, which signals a magnetic response according to Eqs\ (\ref{equation:SFA}) and (\ref{equation:NSFA}). The fit looks even like a simple cosine square function for SF($\theta$)  and a sine square  for NSF($\theta$)  with the same amplitude, which signals $M_y^2=M_x^2$ (see Eqs.\ (\ref{equation:SFA_cosine}) and (\ref{equation:NSFA_sine})).   
The fit parameters $M_y^2$ and $M_x^2$ evaluated at a specific $Q$ range from $SF(\theta)$ and $NSF(\theta)$, are presented in  Fig.\ \ref{fig:x10MxMy} for different temperatures and fields.  
It is obvious in the left panels (a) and (b) that both, the transverse $M_y^2$ and longitudinal $M_x^2$ component, are the same at $T_c$ and stay the same for all $T$ in small fields.

For low fields (of 7 mT) the magnetic response is isotropic in this medium $Q$-regime for high temperatures $T>T_c$ like expected in the PM regime.
The signal is strongest at the critical temperature $T_c$ and decreases for temperatures  above and below $T_c$ as seen in Fig.\ \ref{fig:SFA10} and Fig.\ \ref{fig:x10MxMy}. Note that at the lowest temperature, $T=3$ K, the magnetic signal does not completely vanish. In the FM regime below $T_c$ we need to apply a slightly higher field of 50 mT to keep the neutron beam sufficiently polarized with $P_S>85\%$ for a polarization analysis (see Appendix \ref{sec:depol}). In this small field of 50 mT the response looks still isotropic.
The simple cosine-square or sine-square form with reduced but finite amplitude indicates that isotropic magnetic fluctuations remain in the FM state as observed in the PM state. These magnetic fluctuations change in higher magnetic fields. The $M_y^2$ component in the SF response is reduced while the longitudinal component $M_x^2$ seems less affected. The $M_x^2$ component is more difficult to determine, it is a small disturbance to the $SF(\theta)$ shape or is the main parameter from the $NSF(\theta)$ data that include a high nuclear response.
Through these angle dependent data we are able to confirm magnetic scattering at finite $Q$ in Ni$_{0.90}$V$_{0.10}$ at low $T$. Since they display the same isotropy as in the PM state, it is likely that they are fluctuations with sufficient long time scales to be observed as static by the neutrons. This magnetic signal looks promising to uncover the ``magnetic clusters", the distribution of clusters with different sizes and fluctuation rates that evolve at a disordered QCP. 
 
\begin{figure}
\includegraphics[width=\columnwidth]{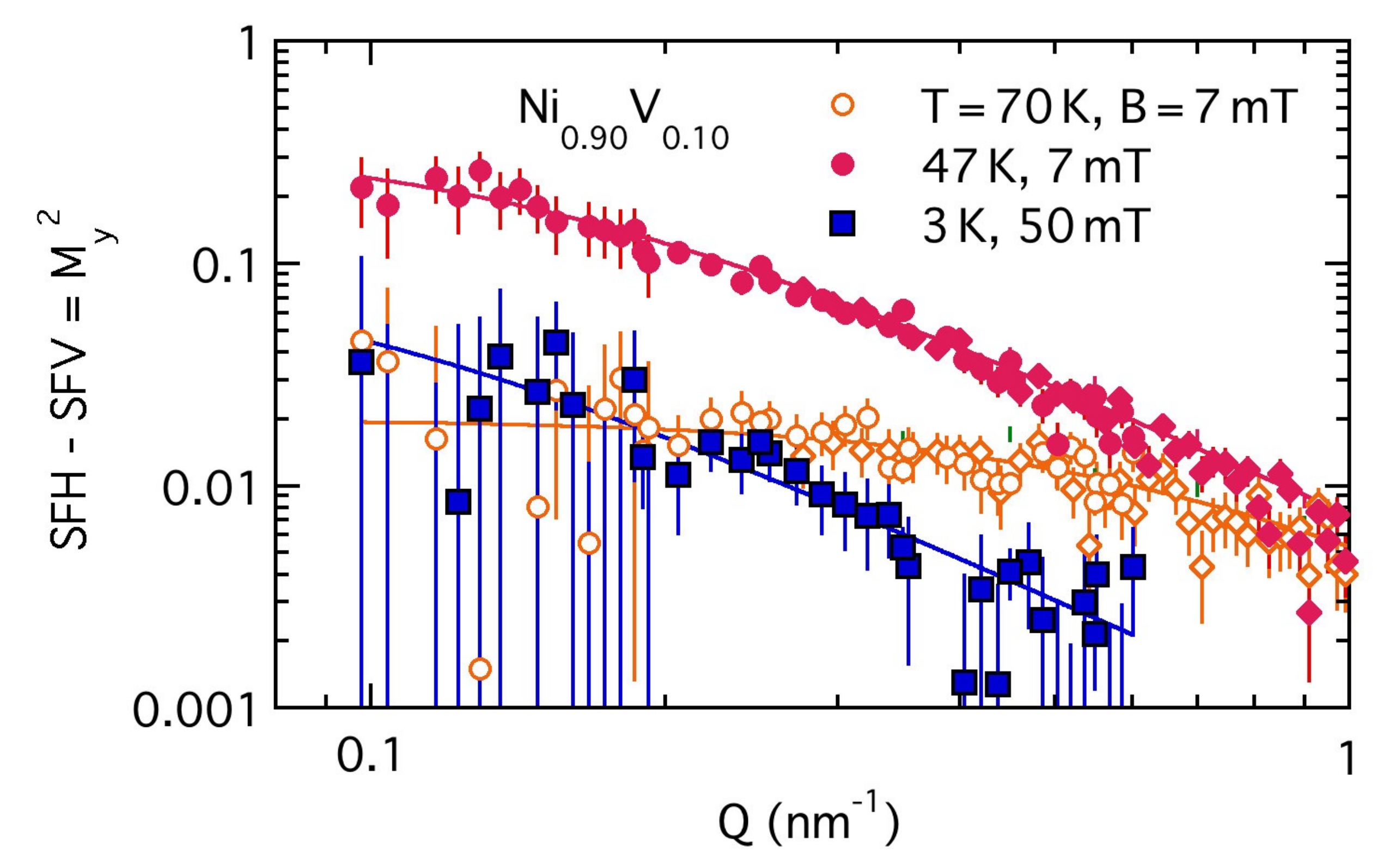}% SFQ10
\caption{\label{fig:both}Magnetic neutron scattering intensity vs. wave vector $Q$ for Ni$_{0.90}$V$_{0.10}$. The spin flip contrast SFH-SFV  is shown in small magnetic fields ($B\leq50$ mT) representing $M_y^2$ or 1/3 of the total isotropic magnetic response $M_{tot}^2$ for different temperatures $T$. Solid lines are Lorentzian fits using Eq.\ (\ref{equation:Lor}).
}
\label{fig:SFQ10}
\end{figure}

%Qdep 
The $Q$-dependence of SANS reveals some direct information about the relevant magnetic correlation length of these clusters. Fig.\ \ref{fig:SFQ10} displays the $Q$-dependence of the SF signal, i.e. the $M_y^2$ component from the SF contrast (SFH-SFV). Since the signal is isotropic (in low fields) $M_y^2$ represents 1/3 of the total magnetic signal $M^2_{tot}$.
$M_y^2$ can be resolved in a limited $Q$-range above 0.1 nm$^{-1}$. Toward lower $Q$, the dominating nuclear NSF contribution increases and makes the extraction of a small SF contribution with polarization corrections less reliable.  The even smaller SF contrast cannot be resolved. The strongest response is found at 47 K close to $T_c$. $M_y^2$ is clearly $Q$ dependent, steeper at $T_c$ than at higher temperatures $T$=70 K. At the lowest temperature of $T=3$ K $M_y^2(Q)$ can be barely resolved beyond the middle-$Q$ region but its $Q$ dependence looks similar to the $T=47$ K data. Within this limited $Q$ region of (0.1-1) nm$^{-1}$ the SF contrast
can be approximated well by a Lorentzian function as expected for paramagnetic critical scattering of the Ornstein-Zernike form with a correlation length $\xi=1/\kappa$: 
\begin{equation}
M_y^2=A_L \times \kappa^2 /(\kappa^2+Q^2).
\label{equation:Lor}
\end{equation}
We estimate a correlation length using Eq.\ (\ref{equation:Lor}) as $\xi\approx (7\pm2)$ nm for $T\approx T_c=47$ K, and we find the comparable value of $\xi=(10\pm6)$ nm in a more restricted $Q$-regime for $T = 3$ K. About $10\%$ of the amplitude $A_L$ of the Lorentzian form remains at low $T$: $A_L(T = 3 \textrm{ K})\approx (1/10) A_L(T = 47 \textrm{ K})$. That corresponds to the same fraction of $M_y^2$ at medium $Q$. 
%It is not uncommon for disordered magnets to observe a maximum signal at finite $Q$ at $T_c$ with a limited correlation length, the nature of the low T excitations depend on the disorder.???\cite{Rhyne1986}
These data clearly show some leftover short-range magnetic correlations in the FM state at $T<T_c$ that do not contribute to the long-range order. The remaining magnetic response at low temperatures is similar to the PM scattering. The cluster sizes or range of correlation length at low $T$ look like those estimates in the PM regime close to $T_c$. The non-polarized data estimates for $x=0.10$ using high field data as reference agree with the main PASANS findings.\cite{Schroeder2020}\\
\begin{figure}
\includegraphics[width=\columnwidth]{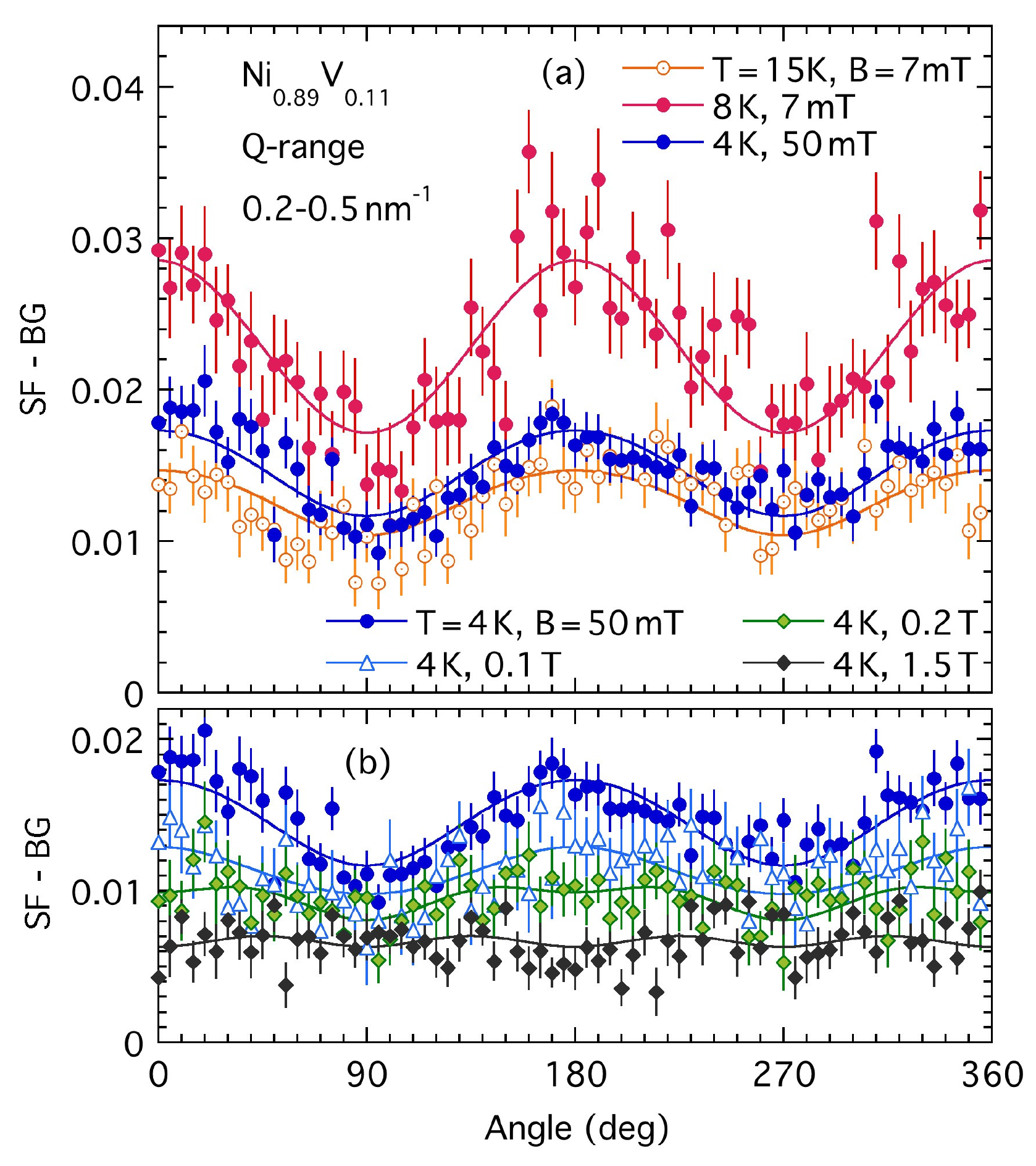}% SFA11
\caption{ Angle dependence of  SF signal collected in medium $Q$ range ($0.35\pm0.15$) nm$^{-1}$ 
for Ni$_{0.89}$V$_{0.11}$. Panel (a) shows data in small magnetic fields $B$ at different temperatures $T$, and panel (b) data at low $T$ in different $B$. 
Fit is shown as solid line using Eq.\ (\ref{equation:SFA}). A constant BG has been subtracted from the data that depends on polarization filter and magnetic field.}
\label{fig:SFA11}
\end{figure}
\begin{figure}
\includegraphics[width=\columnwidth]{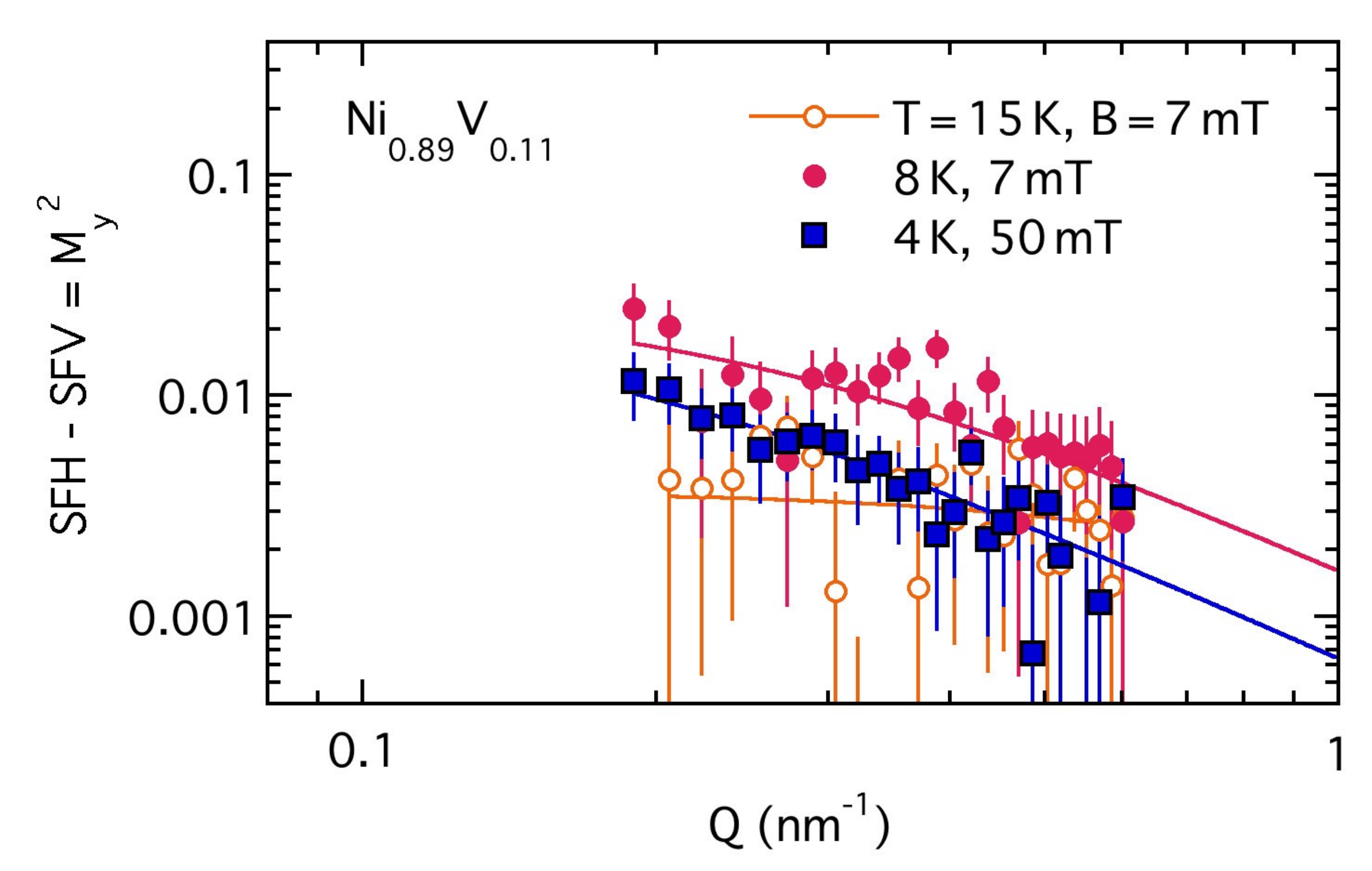}% SFQ11
\caption{\label{fig:both}Magnetic neutron scattering intensity vs. wave vector $Q$ for Ni$_{0.89}$V$_{0.11}$. 
The spin flip contrast SFH-SFV  is shown in small magnetic fields ($B\leq50$ mT) representing $M_y^2$ or 1/3 of the total isotropic magnetic response $M_{tot}^2$ for different temperatures $T$. Solid lines are Lorentzian fits using Eq.\ (\ref{equation:Lor}).
}
\label{fig:SFQ11}
\end{figure}
\begin{figure}
\includegraphics[width=\columnwidth]{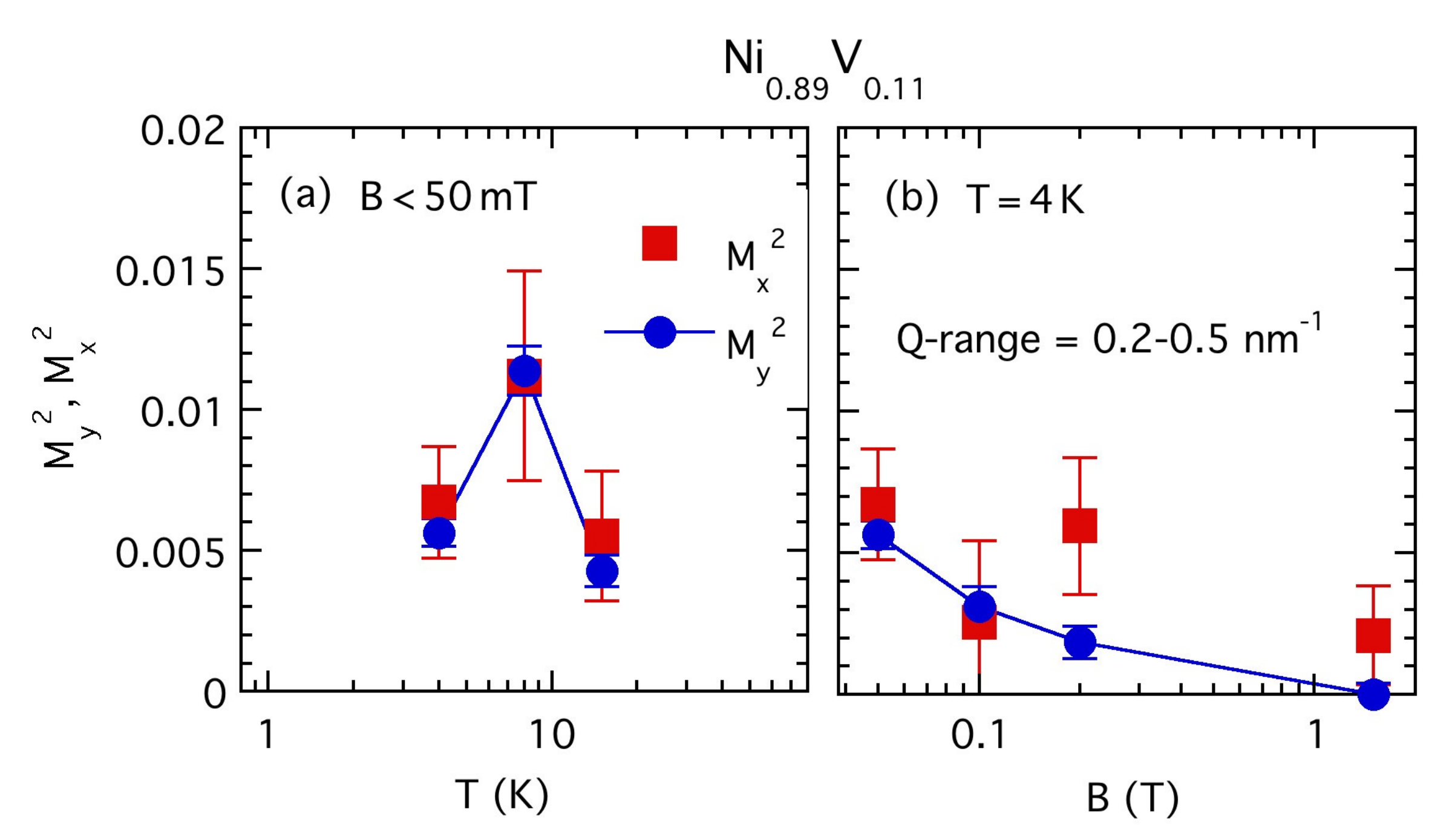}% SFpara11 called SF_fit11 now x11My
\caption{ Magnetic components $M_y^2$ and $M_x^2$ extracted from fit (Eq.\ (\ref{equation:SFA})) of SF($\theta$) data  
(shown in Fig. \ref{fig:SFA11}) 
of Ni$_{0.89}$V$_{0.11}$ with $T_c\approx7$ K at medium $Q$ range of (0.2-0.5) nm$^{-1}$. Panel (a) shows $T$ dependence of isotropic fluctuations in small fields ($B\leq50$ mT) and (b) shows $B$ dependence of short-range correlations at low $T=4$ K, which are different for transverse and longitudinal components. Lines just connect data points. }
\label{fig:x11My}
\end{figure}

\subsection{\label{sec:SF11} Short-Range Correlations in Ni$_{0.89}$V$_{0.11}$}
The discovery of a remaining magnetic signal in Ni$_{0.90}$V$_{0.10}$ motivates us to investigate the more diluted Ni$_{0.89}$V$_{0.11}$ that is closer to critical concentration $x_c$ with a lower $T_c=7$ K. As expected  the magnetic response in Ni$_{0.89}$V$_{0.11}$ is  even smaller than in Ni$_{0.90}$V$_{0.10}$ and more challenging to resolve.
Note that the data shown are not calibrated and a direct comparison of the scattering intensities of $x=0.11$ and $x=0.10$ cannot be made. 
We cannot extract any clear magnetic signal from the NSF data due to the strong nuclear contribution in $x=0.11$. But the  SF data reveal successfully a magnetic response from the distinct angle dependence in a limited $Q$-regime. 
The strongest SF response in Ni$_{0.89}$V$_{0.11}$ close to $T_c$ at $T=8$ K follows a simple cosine square function as seen in Fig.\ \ref{fig:SFA11}(a). All $SF(Q)$ data  can be described well by Eq.\ (\ref{equation:SFA}) confirming an isotropic magnetic response with $M_y^2$=$M_x^2$ in low fields ($B\leq 50$ mT). The $T$ dependence of these transverse and longitudinal components is shown in Fig.\ \ref{fig:x11My}(a). Similarly to Ni$_{0.90}$V$_{0.10}$ the strongest PM response of fluctuating short-range correlations develops upon cooling toward $T_c$, decays for $T<T_c$ but does not vanish for the lowest $T$.  Also the $Q$ dependence of the SF contrast (SFH-SFV) that measures the transverse component $M_y^2$ of the isotropic fluctuations looks similar for both concentrations as displayed in Fig.\ \ref{fig:SFQ10} and \ref{fig:SFQ11}. For Ni$_{0.89}$V$_{0.11}$ the $Q$ range and precision of the data are more limited. $M_y^2(Q)$ can be described by a Lorentzian line shape with Eq.\ (\ref{equation:Lor}). Cooling toward $T_c$ increases the correlation length $\xi=1/\kappa$, but it remains finite at $T_c$. $M_y^2(Q)$ shows about the same $Q$ dependence at the lowest $T$ of 4 K compared to $T_c$. Only the amplitude $A_L$ is reduced.
Fitting the SF contrast at 8 K and 4 K including data up to 1nm$^{-1}$ (not all data shown) yields $\xi=(4\pm2)$ nm and $\xi=(6\pm3)$ nm, respectively.
Within this large uncertainty we conclude that the typical correlation length at $T_c$ and at low $T$ is quite similar in the range of 5-10 nm and not very distinct from Ni$_{0.90}$V$_{0.10}$. These simple estimates show some indication that the magnetic length scale describing the cluster distribution might become smaller towards the critical point, but only better precision estimates from a larger $Q$ regime with improved statistics of the low intensity $x=0.11$ data can clarify that. Most importantly, these polarized data are able to confirm that magnetic clusters are still present at low $T$, which look similar to PM scattering with a typical scale of a few nm. In Ni$_{0.89}$V$_{0.11}$ the magnetic cluster response $M_y^2$  at 4 K is about half of $M_y^2$ at $T_c$ for this medium $Q$ range. 
While the remaining cluster contribution is about of similar magnitude in both samples,  
 the fraction of remaining clusters is much higher in $x= 0.11$ than in $x=0.10$. 
The present data (not shown in the same scale) do not allow to trace a detailed concentration dependence of the remaining clusters. However, we can rule out similar short-range correlations in polycrystalline Ni samples 
(see in Section \ref{sec:Ni0}). 
%Any attempt to find such cluster contribution in pure Ni failed. Within the resolution and the constraints of the data we can set an upper limit that is below the observed values of the alloy with $x=0.10$ (see more in chapter ???IV A)

%
\begin{figure}
\includegraphics[width=.8\columnwidth]{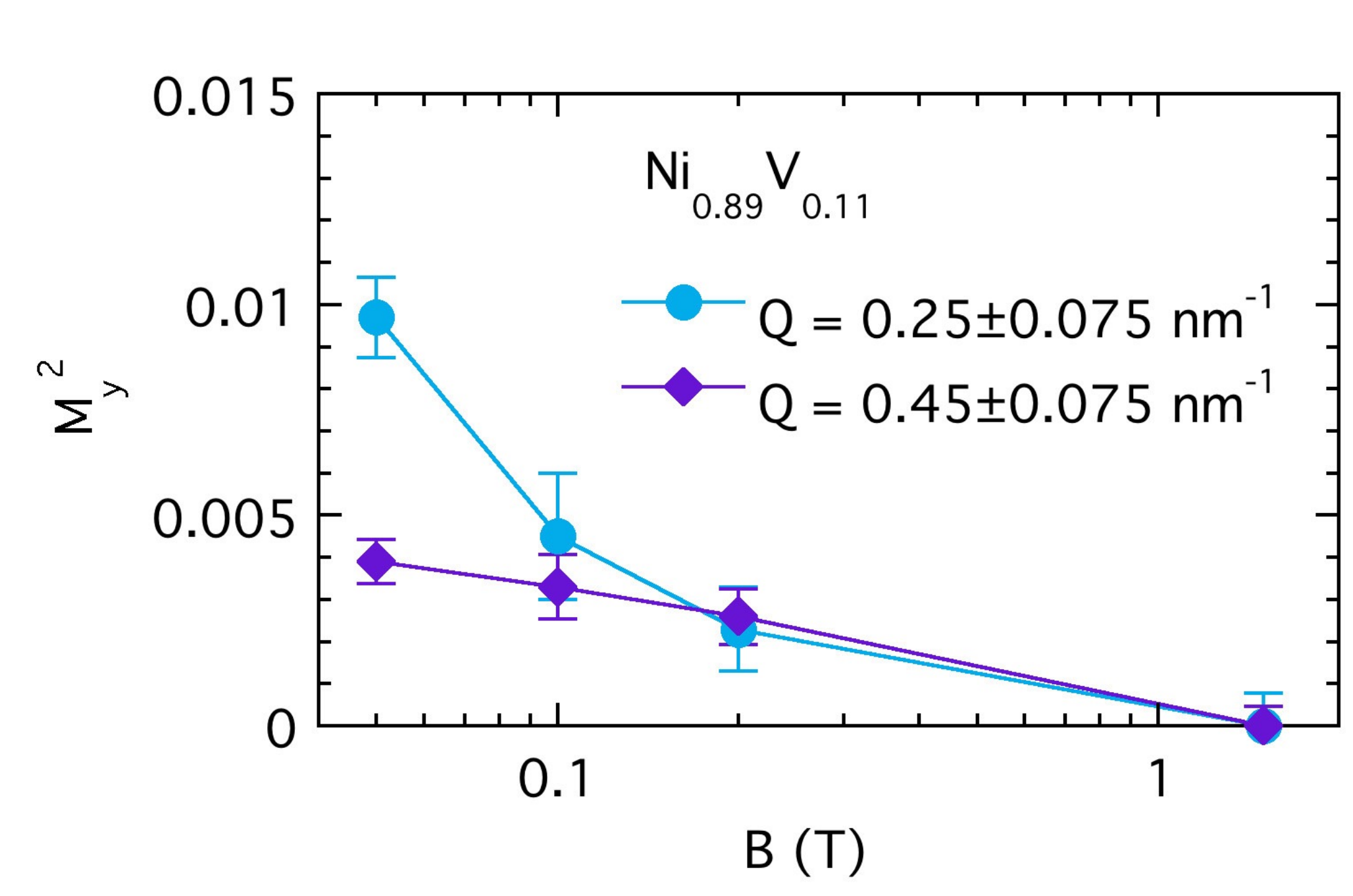}% SF11BdepQ2 other Q range
\caption{ Field dependence of magnetic component $M_y^2$ of Ni$_{0.89}$V$_{0.11}$ at low $T=4$ K. $M_y^2$ values estimated from SF($\theta$) using  Eq.\ (\ref{equation:SFA}) at two different $Q$ ranges as indicated. Lines just connect data.}
\label{fig:x11My2Q}
\end{figure}
%

%In small magnetic fields the magnetic signal remains isotropic even below $T_c$ down to low $T$. 
The signature of a quantum Griffiths phase with a cluster distribution in FM Ni$_{1-x}$V$_x$ came from the field dependence of the bulk magnetization $M$.\cite{Wang2017} 
The characteristic $B$ dependence of $M(B)$ in form of a power law 
indicates that these fluctuations of different sizes and rates are gradually freezing out toward higher $B$. 
%Do these SANS data support this idea and contribute more detailed information? 
The SANS data at low fields confirm already the existence of isotropic magnetic fluctuations at base $T$,
SF data in higher magnetic fields should reveal more detailed information about the distribution. 
We notice in Fig.\ \ref{fig:SFA11}(b) already that the SF response and SF contrast $M_y^2$ at 4 K decrease in higher fields.
The angle dependence also changes, it shows deviations from a pure sine-square form in higher $B$ that indicates an anisotropic magnetic signal. The fit of Eq.\ (\ref{equation:SFA}) reveals different $M_y^2$ and $M_x^2$ components as presented in Fig.\ \ref{fig:x11My}(b). The transverse component $M_y^2$ is decreasing with $B$ while the longitudinal component $M_x^2$ does not change within  large error bars. The Ni$_{0.90}$V$_{0.10}$ data (Fig.\ \ref{fig:SFA10} (a,b), Fig.\ \ref{fig:x10MxMy} (c,d)) support the same trend: $M_y^2$ decreases with $B$ while $M_x^2$ remains finite. Both SF($\theta$) and NSF($\theta$) yield consistently a dominant $M_x^2$ component in high field. 
We did not expect to see a remaining magnetic longitudinal component $M_x^2$  at finite $Q$ in strong magnetic fields. Since these SF data do not allow to explore further the $Q$ dependence of this term we postpone the discussion to the  the next sections. In Secs. \ref{sec:DIF10} and \ref{sec:DIF11} we employ another method DIF that is better suited to resolve this anisotropic term.
 
We do understand and expect that $M_y^2$ decreases with $B$. That signals that the fluctuations indeed freeze out in higher field, more moments from short-range correlations do not contribute to this higher $Q$ region since they might join larger domains. 
 %We check later in the next chapter how to access the domain contribution and field dependance. 
Unfortunately, we cannot resolve the detailed $Q$ dependence and $B$ dependence to find a specific cluster distribution. 
The statistics are too poor to even extract a Lorentzian fit with two parameters from $M_y^2(Q)$ in high fields that is further reduced compared to low fields.  
Fitting the angle dependence  of SF data collected in wide $Q$-ranges of $\Delta Q=0.15$ nm$^{-1}$ provides sufficient statistics to trace $M_y^2(B)$ at two different $Q$ locations.  
Fig.\ \ref{fig:x11My2Q}  presents 
 $M_y^2(B)$ for Ni$_{0.89}$V$_{0.11}$ obtained from $SF(\theta)$ at different $Q$, comparing a lower $Q$ range  (centered at $0.25$ nm$^{-1}$) and  a higher $Q$ range (centered at $0.45$ nm$^{-1}$). 
We notice that $M_y^2$ shows a steeper $B$-dependence at lower $Q$ than at higher $Q$. This trend is consistent with the idea that larger slower clusters freeze out in smaller fields than smaller clusters with higher fluctuations rates. Our data might support a distribution of fluctuating cluster sizes consistent with quantum Griffiths singularities. Better resolution is certainly necessary to see and confirm more details.

%%%%%%%%%%%%%%%%%%%%%%%%%%%%%%%%%%%%%%%%%%%%%%%%%%%%%%%%%%%%%%%%%%%%%%%%%%%%%%%%%%%%%%%%%%%%%%%%%%%%%
%DIF
%%%%%%%%%%%%%%%%%%%%%%%%%%%%%%%%%%%%%%%%%%%%%%%%%%%%%%%%%%%%%%%%%%%%%%%%%%%%%%%%%%%%%%%%%%%%%%%%%%%%%
\subsection{\label{sec:DIF10}Search for Long-Range Order in Ni$_{0.90}$V$_{0.10}$}
Another way to extract the longitudinal magnetic scattering is to consider the difference response “DIF” between the two initial polarization directions without analyzing the neutron polarization after the sample. The NSF asymmetry or flipper difference is collected from fully polarized data (DD-UU) or half polarized (HP) data (D-U) as introduced before in Section \ref{sec:pol}, where Eq.\ (\ref{equation:DIF}) presents the characteristic angle dependence of DIF($\theta$).
This interference term of nuclear and magnetic origin, $2NM_x$, more easily confirms a weak magnetic signal with a strong nuclear contribution. It is therefore
accessible 
in a larger $Q$ regime in  Ni$_{1-x}$V$_{x}$ providing further information on different length scales than the fluctuations from the SF signal.
%
%DIFAx10
\begin{figure}
\includegraphics[width=0.8\columnwidth]{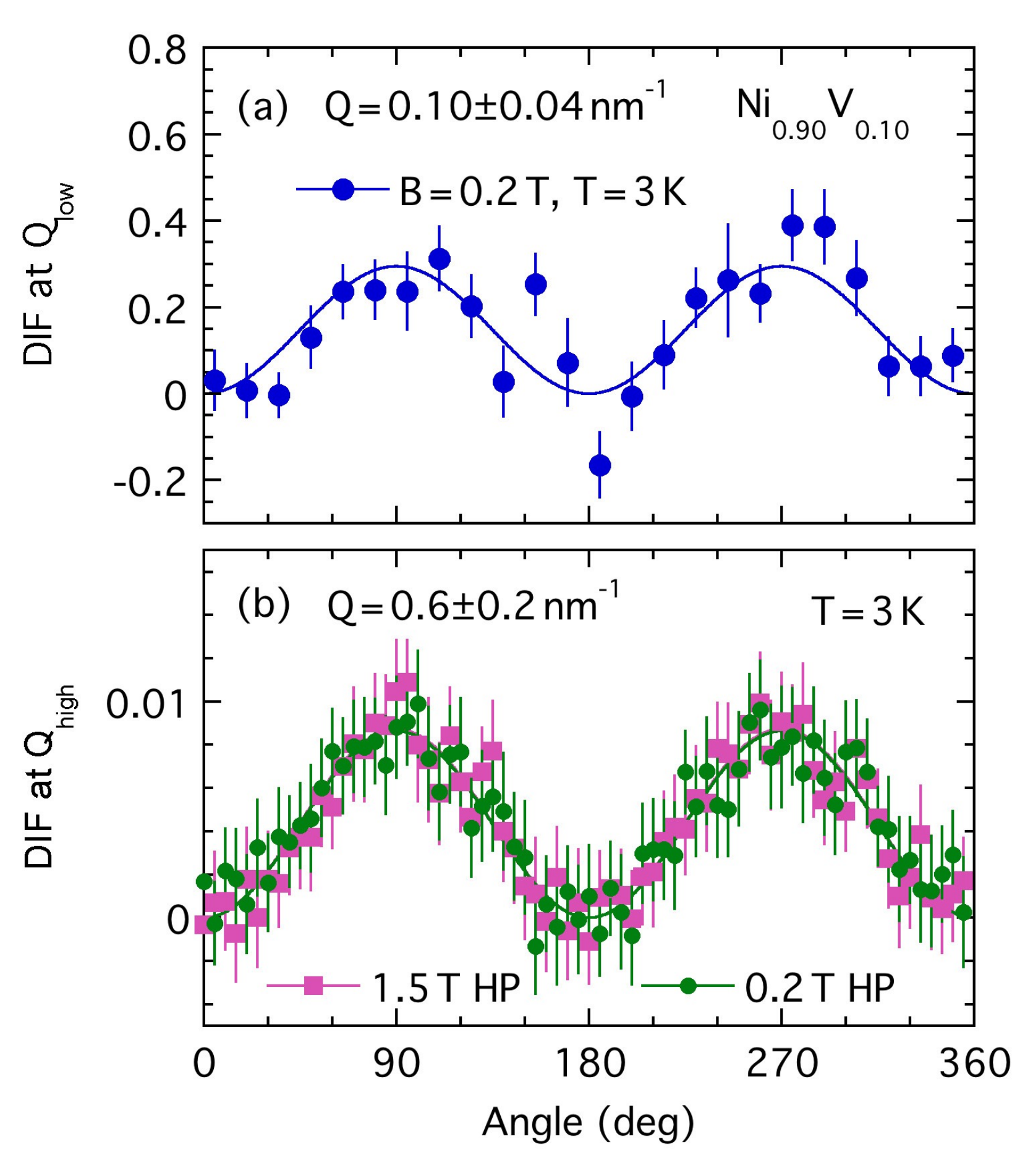} %DIFA10
\caption{Angle dependence of neutron response DIF$(\theta)$ of Ni$_{0.90}$V$_{0.10}$ for different $Q$ ranges at low $T=3$ K $<T_c$ confirming magnetic response $2NM_x$, solid lines follow Eq.\ (\ref{equation:DIF}). (a) Low $Q$ range covers $(0.06-0.14)$ nm$^{-1}$, (b) high $Q$ range includes $(0.4-0.8)$ nm$^{-1}$. HP denotes half polarized data, where DIF=D-U.
}
\label{fig:DIFA10}
\end{figure}
\begin{figure}
\includegraphics[width=0.9\linewidth]{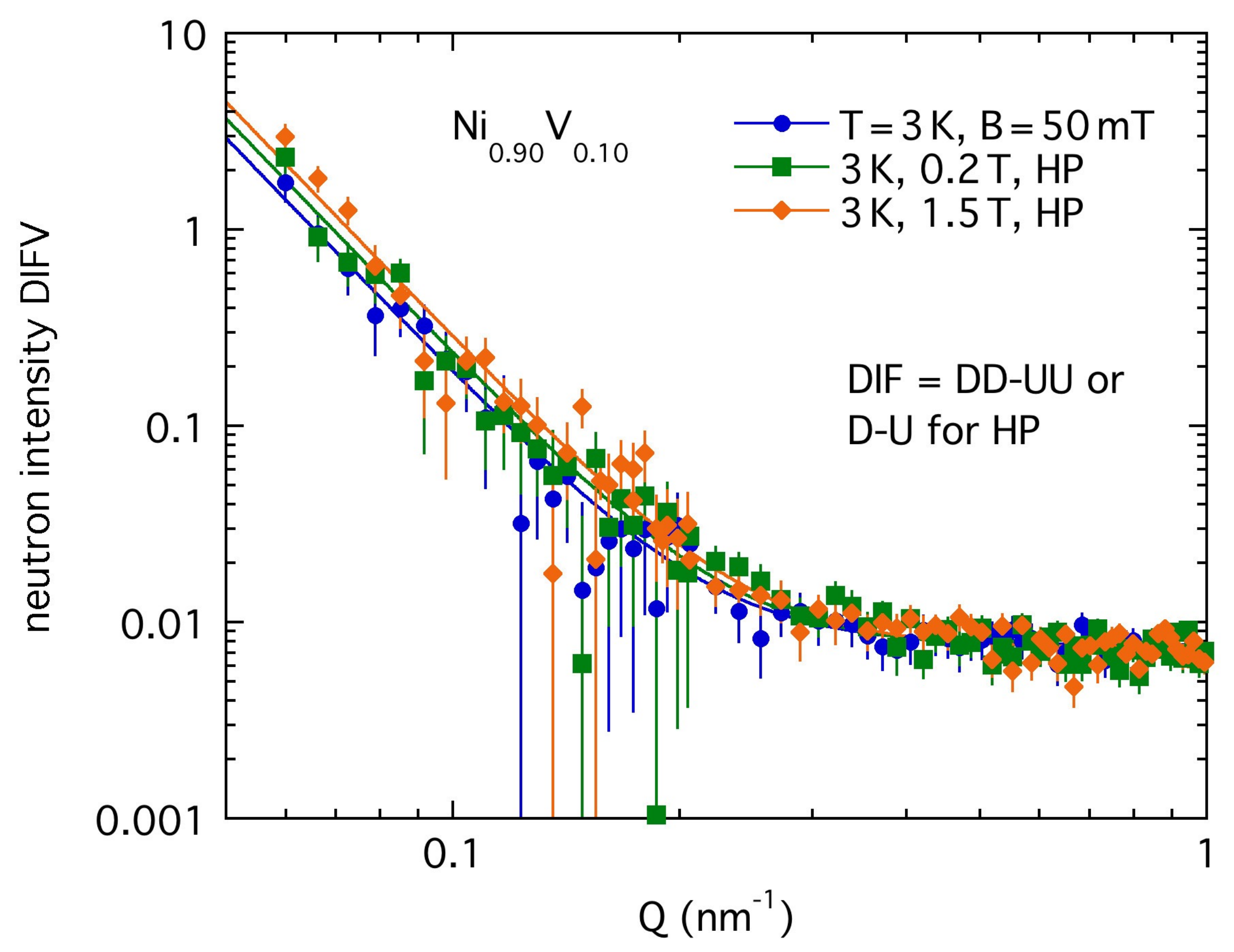} %DIFV10
\caption{$Q$ dependence of DIFV=$2NM_{x}$ of Ni$_{0.90}$V$_{0.10}$ at low $T$ = 3 K $< T_c$ in magnetic fields $B$. Solid line indicates fit according to Eq.\ (\ref{eq:DIFV}).
}
\label{fig:DIFV10}
\end{figure}
\begin{figure}
\includegraphics[width=1\linewidth]{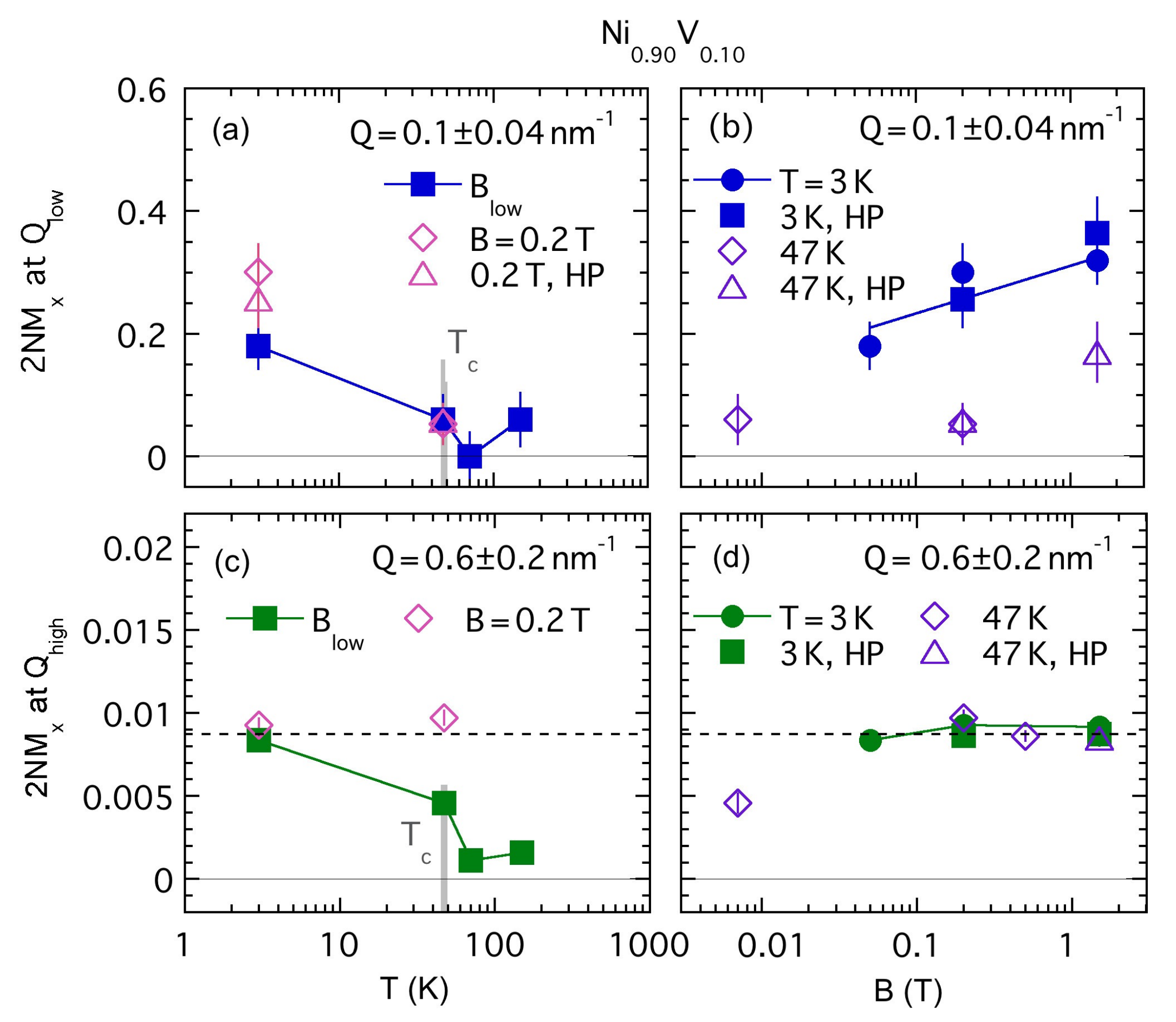} %2NMX10para
\caption {Magnetic response $2NM_x$ of Ni$_{0.90}$V$_{0.10}$ for low $Q$ (upper panels (a,b)) and high $Q$ (lower panels (c,d) derived from DIF($\theta$). The left panels (a,c) compare the temperature dependence in small magnetic fields as indicated ($B_{\textrm{low}}\leq50$ mT), the right panels (b,d) the field dependence at $T=3$ K $<T_c$ and $T=47$ K $=T_c$. While $2NM_x$ at $Q_{\textrm{low}}$ represents the long-range magnetic domain contribution, $2NM_x$ at $Q_{\textrm{high}}$ signals local magnetic impurities. The solid lines connect data points in (a,c,d).
}
\label{2NMX10}
\end{figure}

First we  check the angular dependence of the DIF data in Ni$_{0.90}$V$_{0.10}$ for traces of a magnetic response in different $Q$ regimes. Fig.\ \ref{fig:DIFA10} shows DIF($\theta$) at low $T=3$ K in $B=0.2$ T for a low $Q$ region (centered at $Q_{\textrm{low}}=0.10$ nm$^{-1}$) in panel (a) and a high $Q$ region (centered at $Q_{\textrm{high}}=0.6$ nm$^{-1}$) in panel (b). The DIF$(\theta)$ data follow well the expected response of a sine square function  (see Eq.\ (\ref{equation:DIF})) that confirms a longitudinal magnetic contribution $M_x$ in the magnetized state. The maximum amplitude observed at $DIF(\theta=90^\circ)$=DIFV=$2NM_x$ is the only fit parameter.
% of Eq.\ (\ref{equation:DIF}), is present in the whole $Q$-range, but quite different for these extreme $Q$-values.
Fig.\ \ref{fig:DIFV10} presents the detailed $Q$-dependence of DIFV in a wide $Q$ range from $Q=(0.06-1)$ nm$^{-1}$ for Ni$_{0.90}$V$_{10}$ at base $T=3$ K well below $T_c$. 
As long as the applied field is large enough ($B\geq50$ mT) that the neutron beam does not become depolarized ($P_s >85\%$) in the FM state of the sample, the DIFV($Q$) response looks similar for different $B$. Towards low $Q$, the signal presents a steep $1/Q^n$ upturn that follows a ``Porod" term \cite{Porod} with power $n=4$ . Toward higher $Q$ the intensity remains constant. DIF($Q$) can be represented by 
%a fit (Eq.\ (\ref{eq:DIFV})) with two terms as shown as solid line in Fig.\ \ref{fig:DIFV10}.
%
\begin{equation}
DIFV(Q)=A_D/Q^4 + C_D
\label{eq:DIFV}
\end{equation}
We recognize two different responses of aligned magnetic moments $M_x$ in the $B$ direction dominant at lower $Q$ and remaining at higher $Q$ in the FM state at 3K. To distinguish their origin we study the evolution with $T$ and $B$. 
 
The upper panels of Fig.\ \ref{2NMX10}(a,b) focus on the interference term $2NM_x$ at $Q_{\textrm{low}}$  representative of the low $Q$ upturn with amplitude $A_D$.
The $2NM_x(Q_{\textrm{low}})$ signal is present at the lowest temperature in fields $B\geq 50$mT that can be analyzed in the FM state. As shown in panel (a) a clear positive response of $2NM_x$ appears only in the FM phase. $2NM_x$ becomes very small close to zero toward higher $T$ when crossing the critical temperature $T_c=47$ K into the PM state. The minor deviation from zero at higher $T$ is not significant.  The signal at low $T$ is further increasing by applying a higher $B$ as shown in panel (b). The field also changes the response in the PM regime at $T=47$ K $ \approx T_c$ but much less than in the ordered state below $T_c$. 
Further higher precision data at several $T$ below and above $T_c$ could trace and reveal the onset of a FM response in more detail.
However, these changes in $2NM_x(T,B)$ are clearly related to the ``magnetic" contribution $M_x$ that appears below and vanishes above $T_c$.
  
As this mixed term $2NM_x$ includes the nuclear contribution, we check the pure nuclear contribution $N^2$ separately for any anomalies in $T$, $B$ and in particular in $Q$ to get a better estimate of the pure magnetic response.   
 The pure nuclear signal from coherent scattering from Ni, $N^2$, is estimated from the total NSF contribution in a sector along the field direction ($NSFH = N^2+BG$) after other BG subtraction of mainly sample holder. $N^2(Q)$ shows a $1/Q^4$ dependence toward low $Q$ (see Ref. \onlinecite{Schroeder2020}) without any obvious $T$ or $B$ dependence. This strong nuclear response in this lower $Q$ range stems from grain boundaries of crystallites on the order of $\mu m$ in these polycrystalline samples. The grain size is too large to be determined from these SANS data in this $Q$ regime ($1\mu$m $> 1/Q_{min} = 1/0.05 $ nm$^{-1}$). 
 Since the nuclear $N^2(Q)$ and the cross term $2NM_x(Q)$ exhibit a $1/Q^4$ dependence in the observed $Q$ range, the   magnetic term $(M_x)^2(Q)$ follows then also a $1/Q^4$ dependence from the simple estimate of $(M_x)^2=(2NM_x)^2/(4N^2) \sim Q^{-4\times2}/Q^{-4}=Q^{-4}$. The precision of $2NM_x$ and $M_x$ is far less than $N^2$ but deviations from $1/Q^4$ are not obvious. Therefore the $Q$ dependence of $2NM_x(Q)$ does not contradict and rather might support large-scale magnetic domains of similar order to the grain sizes. This is in agreement with the simple depolarization estimates (see Appendix \ref{sec:depol}). 
 The fact that the sample depolarizes the neutron beam strongly indicates FM order of large-scale domains on the order of  $\mu m$.
 Lower $Q$ data with better precision might increase the chance to notice deviations from power $n=4$. This could reveal indication of a fractal nature of perforated domains \cite{Kreyssig2009} or reduced domain sizes. Recognizing  saturation effects at lower $Q$ might improve estimates of a domain size, e.g. assuming a simple Lorentzian square fit \cite{Hellman1999}. So far, the lower limit is only 50nm (see Ref. \onlinecite{Schroeder2020}).
 Note that e.g. nanocrystalline Ni with average crystallite size of about 50nm presents a different response \cite{Weissm2001}. DIF(Q) is weakly $Q$-dependent \cite{Weissm2001}, it does not display the steep $1/Q^4$ upturn, or the constant $Q$ term toward higher $Q$.

This DIFV signal at low $Q$ in Ni$_{0.90}$V$_{0.10}$ shows the evolution of magnetic domains, long-range ordered regions, that develop below $T_c$. As we saw in the previous Section\ \ref{sec:SF10} some short-range fluctuations remain at low $T$, but most of the Ni moments seem to form still a long-range ordered network in the FM state below $T_c$. We do not have evidence that the overall macroscopic domain size is reduced to short-range order or cluster freezing by the introduction of disorder through V in this alloy. 
As we see in Fig.\ \ref{fig:DIFV10} and Fig.\ \ref{2NMX10} the upturn $2NM_x(Q_{\textrm{low}})$ still increases with $B$ at low $T=3$ K. At the same fields we notice that $M_y^2$ at higher $Q$ range from SF is decreasing. While the short-range cluster fluctuations are freezing out, we expect that the contribution to the long-range magnetic domains are growing. Indeed, the data support consistently that the magnetic clusters freeze out gradually to join the ordered net moments that increase in field direction.  
Since the nuclear response does not show any anomaly at a magnetic transition, the response of $2NM_x$($T,Q_{\textrm{low}}$) at low $B$ marks the onset of FM order. 
%More quantitative comparison of M$_x$ between different samples will be discussed later.

%[hiQ]
The second magnetic term in $2NM_x(Q)$ that becomes dominant at higher $Q>0.2$ nm$^{-1}$ is rather $Q$ independent within the investigated $Q$-regime up to 1nm$^{-1}$. The $T$ and $B$ dependence is different for this ``high" $Q$ term and points to a different origin than the low $Q$ term. This constant $C_D$ or $2NM_x(Q_{\textrm{high}})$ gradually decreases with increasing $T$ without vanishing at $T_c$ as shown in panel (c) of Fig.\ \ref{2NMX10}. 
Also, $2NM_x(Q_{\textrm{high}})$ does not change much with higher magnetic fields (see panel (d)). It rather saturates at $2NM_x^{max}$ already in $B\geq50$ mT for low and higher $T=T_c$, different 
than the low $Q$ upturn that was related to long-range magnetic domains. The $Q$-independence supports that this high $Q$ 2NM$_x$ contribution stems from rather local defects in this Ni$_{0.90}$V$_{0.10}$ alloy. An extended $Q$-range to larger $Q$ is essential to probe the effective length scale of these defects.  They become visible when the material is magnetized through entering the FM state forming domains or through an external magnetic field. We believe that this is not an instrumental artifact but a signature of defects introduced by the vanadium as we will show later in Section \ref{sec:Ni0} through comparison with pure Ni data.\\

\subsection{\label{sec:DIF11}Search for Long-Range Order in Ni$_{0.89}$V$_{0.11}$}

\begin{figure}
\includegraphics[width=0.8\columnwidth]{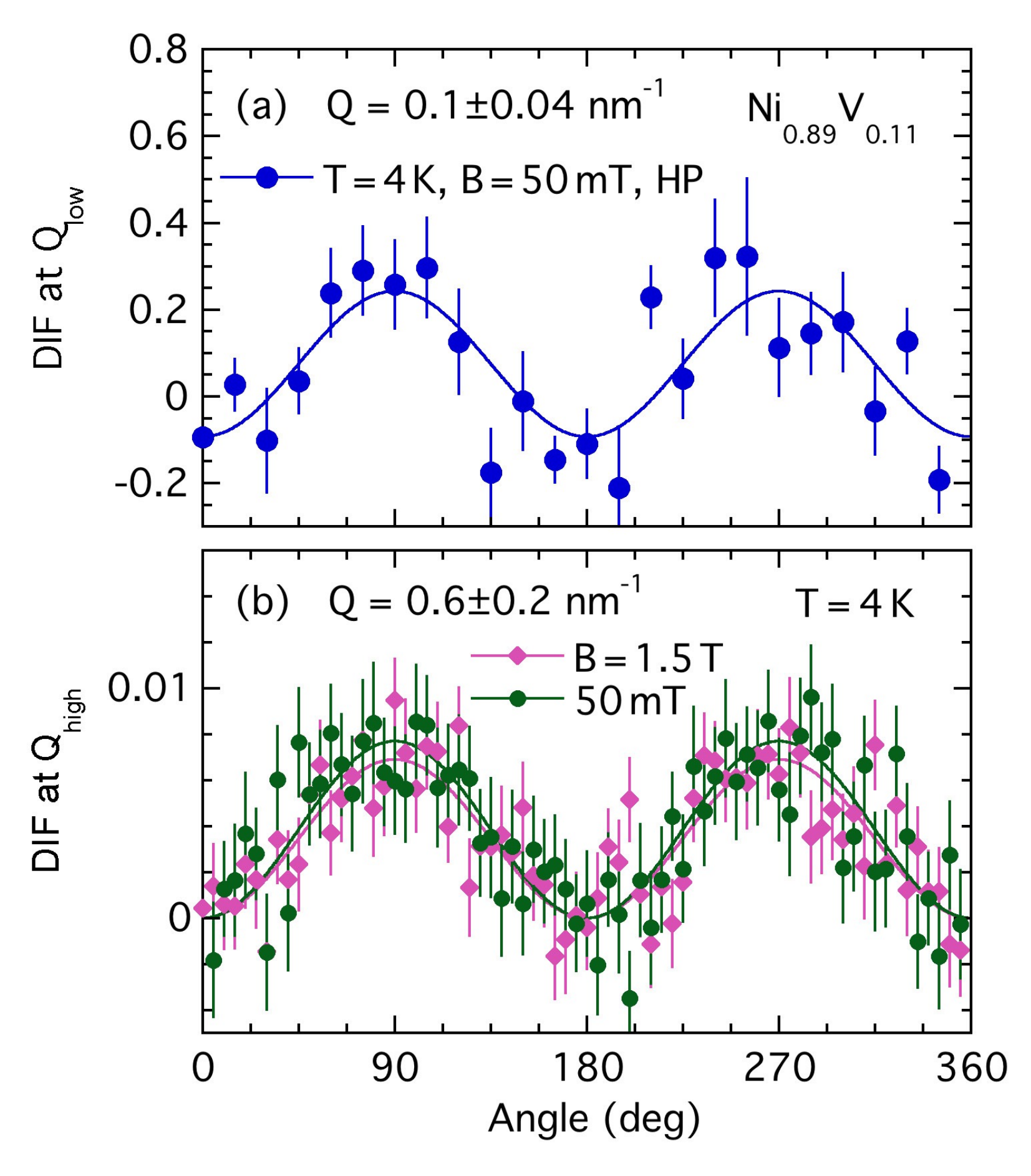}%DIFA11
\caption{Angle dependence of DIF$(\theta)$ of Ni$_{0.89}$V$_{0.11}$ for different $Q$-ranges at low $T=4$ K $ <T_c$ confirming magnetic response $2NM_x$; solid lines follow Eq.\ (\ref{equation:DIF}) with a small offset. (a) Low $Q$ range covers (0.06-0.14) nm$^{-1}$, (b) high $Q$ range includes (0.4-0.8) nm$^{-1}$. HP denotes half polarized data, where DIF=D-U. 
}
\label{fig:DIFA11}
\end{figure}

\begin{figure}
\includegraphics[width=0.9\columnwidth]{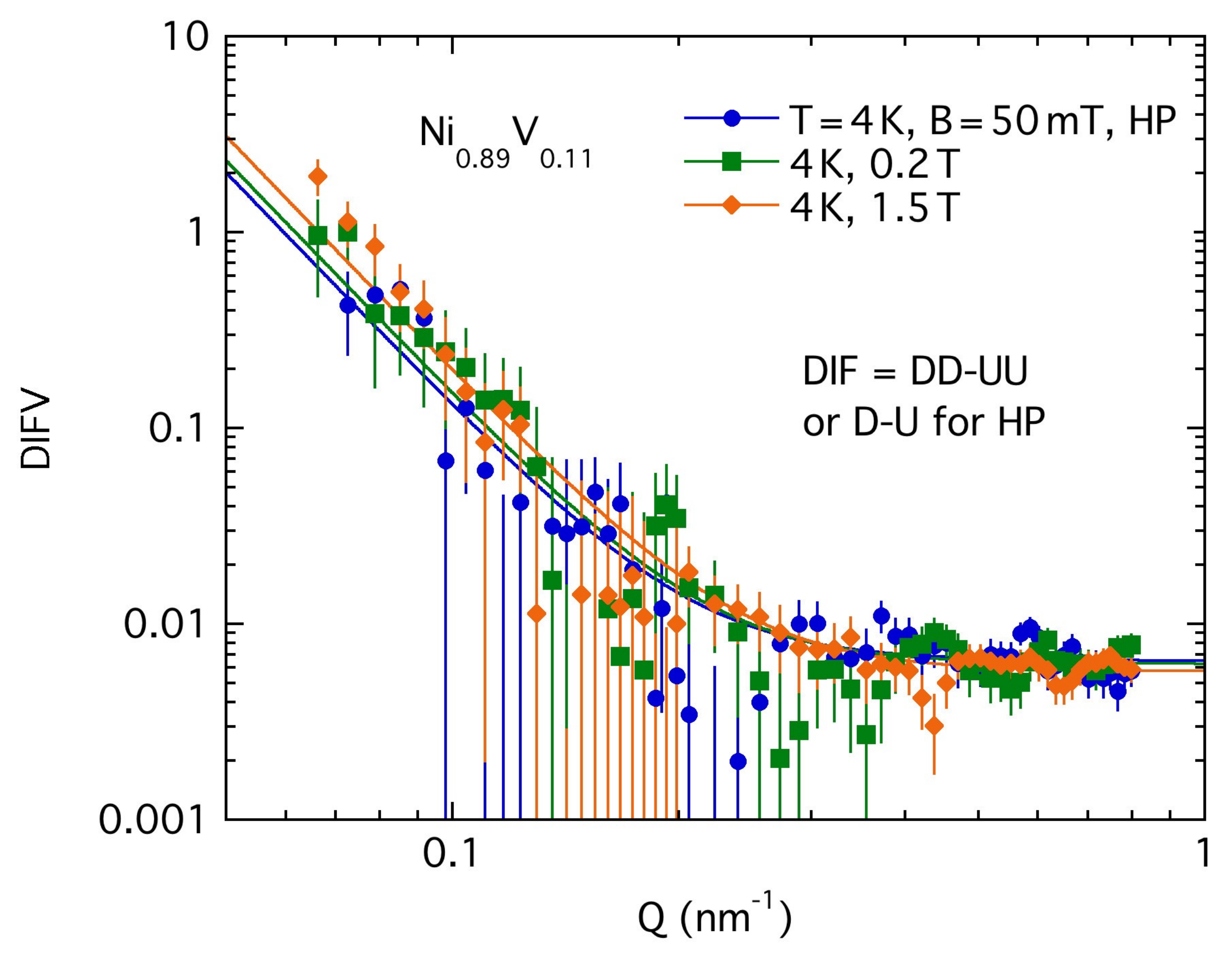}% DIFV11
\caption{ $Q$ dependence of DIFV=$2NM_{x}$ of Ni$_{0.89}$V$_{0.11}$ at low $T = 4$ K $< T_c$ in magnetic fields $B$. Solid line indicates fit according to Eq.\ (\ref{eq:DIFV}).
}
\label{fig:DIFV11}
\end{figure}

Next, we explore the DIF response for the other concentration $x=0.11$ closer to $x_c$. Also here we find signs of a longitudinal magnetic response $M_x$ below $T_c$ from the cross term $2NM_x$ in a large $Q$ regime similar to $x=0.10$.
Fig.\ \ref{fig:DIFA11} presents the characteristic angle variation 
  of $\sin^2(\theta)$ in $DIF(\theta)$ (see Eq.(\ref{equation:DIF})) at extreme $Q$ regions, at $Q_{\textrm{low}}=0.10\pm0.04$ nm$^{-1}$ in panel (a) and at $Q_{\textrm{high}}=0.6\pm0.2$ nm$^{-1}$ in panel (b) at the lowest temperature $T=4$ K for $B\geq0.05$ T.
Fig.\ \ref{fig:DIFV11} shows then the $Q$ dependence of the maximum response at DIFV at $T=4$ K for $B\geq50$ mT.
It can be described by Eq.\ (\ref{eq:DIFV}) by a $1/Q^4$ upturn and a remaining constant $C_D$ similar to DIFV($Q$) of Ni$_{0.90}$V$_{0.10}$. 
We see that this low $Q$ scattering related to magnetic domains is increasing slightly with the magnetic field $B$, while the constant at high $Q$ does not show any variation with the field.  This extra DIFV=$2NM_x$ contribution at high $Q$ confirms clearly a net magnetic $M_x$ term that backs up the anisotropic component $M_x^2>M_y^2$ from SF that persists in high fields as presented earlier in Fig.\ \ref{fig:x11My} in Section \ref{sec:SF11}.

\begin{figure}
\includegraphics[width=1\linewidth]{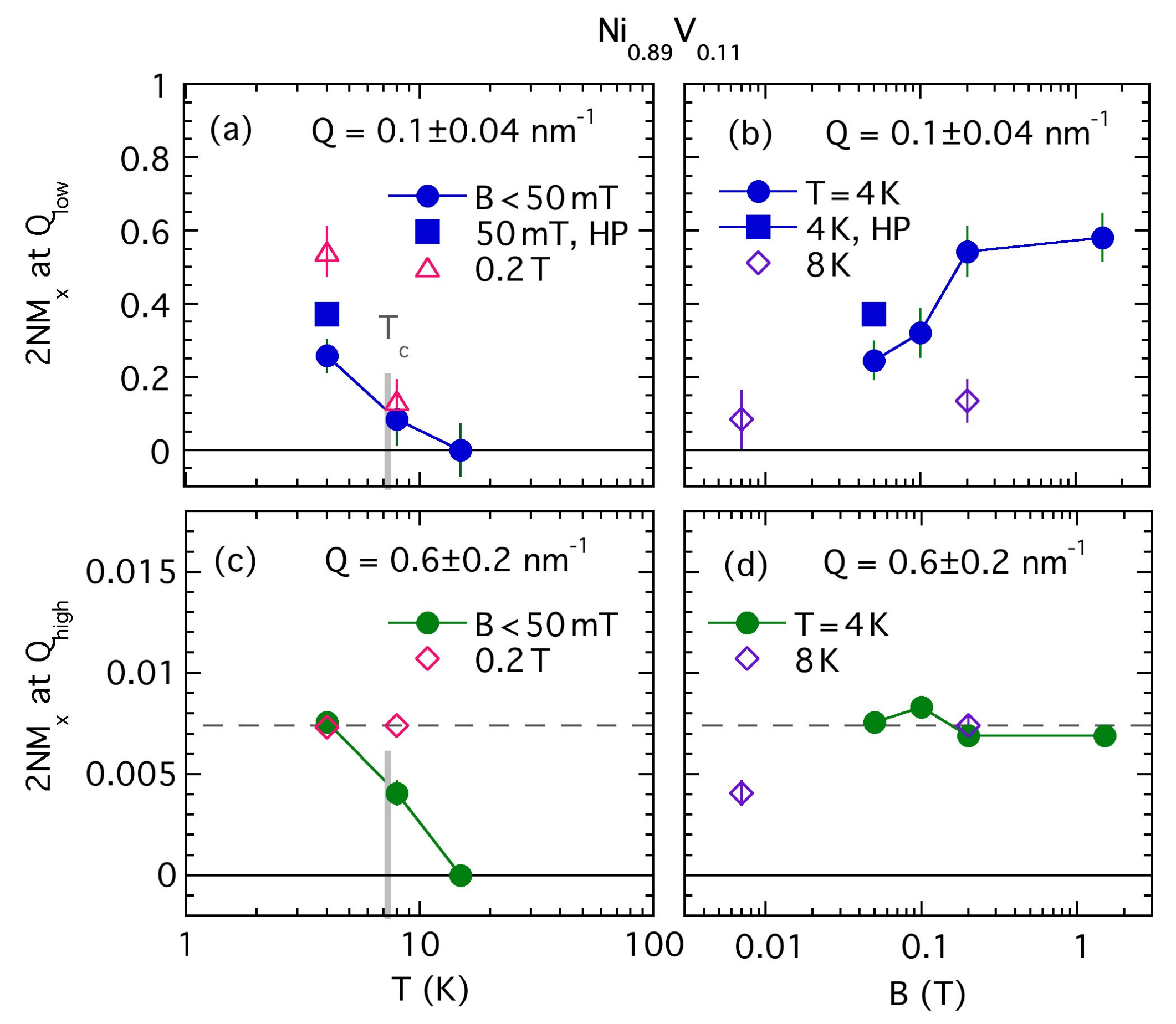}%DIFpara11
\caption{Magnetic response $2NM_x$ of Ni$_{0.89}$V$_{0.11}$ for low $Q$ (upper panels (a,b)) and high $Q$ (lower panels (c,d) derived from DIF($\theta$). The left panels (a,c) compare the temperature dependence in small magnetic fields as indicated ($B_{low}\leq50$ mT), the right panels (b,d) the field dependence at $T=4$ K $<T_c$ and $T=8$ K $\approx T_c$. While $2NM_x(Q_{\textrm{low}})$ represents the ordered domain contribution, $2NM_x(Q_{\textrm{high}})$
stems from local magnetic impurities.
}
\label{fig:2NMX11}
\end{figure}

This cross term DIFV that evaluates $2NM_x$ (derived from the difference of dominant nuclear responses) is of similar scale in $x=0.11$ but shows more overall scatter than for $x=0.10$. This is not surprising, we expect a smaller $M_x$ (comparing $M(B)$ data\cite{Wang2017}) and a larger $N^2$ for Ni$_{0.89}$V$_{0.11}$ than for Ni$_{0.90}$V$_{0.10}$. The $x=0.10$ sample contains ``natural" Ni, while the $x=0.11$ sample is made from pure isotope $^{58}$Ni with twice the coherent cross section of natural Ni. %Nevertheless the angular variation can still be resolved for low Q. 
 For $x=0.11$ we take the maximum variation, the contrast between DIFV and DIFH, as the best $2NM_x$ estimate from $DIF(\theta)$. We accepted consistently for all runs a negative DIFH = -0.18/-0.12 (pol/HP) instead of ideally 0. This  small deviation of  $<1\%$ of the large NSF signal seems within the resolution of the instrument and analysis. The parameter $2NM_x$ is evaluated for $x=0.11$ for several temperatures and fields as shown in Fig.\ \ref{fig:2NMX11} for $Q_{\textrm{low}}$ (upper panels) and $Q_{\textrm{high}}$ (lower panels), similar to Fig.\ \ref{2NMX10} for $x=0.10$.

The low $Q$ upturn in $2NM_x$ is represented through $2NM_x(Q_{\textrm{low}})$ in Fig.\ \ref{fig:2NMX11}(a,b). It is present in the FM state at low $T$ in magnetic fields ($B\geq0.05$ T) that keep the beam sufficiently polarized. Panel (a) confirms that it develops only below $T_c$ and marks the onset of FM order with long-range domains. The ``precise" $Q$-dependence for only $(M_x)^2(Q)$ cannot be revealed from DIFV($Q$), but it does not contradict a $1/Q^4$ dependence without deviations due to long-range magnetic domains as explained for Ni$_{0.90}$V$_{0.10}$ above. Additional support for long-range ferromagnetic order comes from the fact that this Ni$_{0.89}$V$_{0.11}$ sample depolarizes the beam below $T_c$ (see Appendix \ref{sec:depol}). A simple estimate from neutron depolarization  suggests a domain size of few $\mu$m. Data toward lower $Q$ with better statistics are necessary to clarify more details.

The amplitude of the low $Q$ upturn recorded at low $T$, $2NM_x(Q_{\textrm{low}})$, is gradually growing in higher fields as shown in Fig.\ \ref{fig:2NMX11}(b) and Fig.\ \ref{fig:DIFV11}. Previously in Section \ref{sec:SF11} we saw that the fluctuations $M_y^2$ at high $Q$ decrease with $B$. This consistently supports the picture of short-range clusters joining larger domains. Ni$_{0.89}$V$_{0.11}$ shows similar magnetic signatures to  Ni$_{0.90}$V$_{0.10}$, mainly that large-scale magnetic domains still form in the FM state although short-range magnetic fluctuations are becoming more dominant.
The high $Q$ DIF signal or remaining constant $C_D$ represented by $2NM_x(Q_{\textrm{high}})$ is also present in Ni$_{0.89}$V$_{0.11}$. The lower panels (c,d) of Fig.\ \ref{fig:2NMX11} show that it vanishes gradually with increasing temperature and saturates already in medium fields at any $T$.  Similarly to Ni$_{0.90}$V$_{0.10}$ any external field or FM state is sufficient to magnetize the sample to notice these defects in Ni$_{0.89}$V$_{0.11}$.

The more diluted sample Ni$_{0.89}$V$_{0.11}$ follows the same trend of Ni$_{0.90}$V$_{0.10}$, the magnetic $2NM_x$ signal is essentially zero at temperatures above $T\geq T_c$ but appears below $T_c$. This indicates FM order within macroscopic domains even very close to the critical point.    
We collected three different magnetic responses in Ni$_{0.90}$V$_{0.10}$ and Ni$_{0.89}$V$_{0.11}$ through SF and DIF in different $Q$ ranges. For both samples, we could extract short-range fluctuations remaining at low $T$ but also find signatures of long-range domain scattering. To further check the origin and the importance for a disordered alloy we compare the signals with pure Ni and discuss the evolution with the vanadium dilution.

%%%%%%%%%%%%%%%%%%%%%%%%%%%%%%%%%%%%%%%%%%%%%%%%%%%%%%%%%%%%%%%%%%%%%%%%%%%%%%%%%%%%%%%%%%%%%%%%%%%%%
%discussion: evolution with V
%%%%%%%%%%%%%%%%%%%%%%%%%%%%%%%%%%%%%%%%%%%%%%%%%%%%%%%%%%%%%%%%%%%%%%%%%%%%%%%%%%%%%%%%%%%%%%%%%%%%%

\section{\label{sec:discussion} Discussion}
\subsection{\label{sec:Ni0} SANS Response of Ni vs. Ni-V Alloy}
To clarify which magnetic effects are induced by vanadium, we compare the magnetic responses of the alloy Ni$_{0.90}$V$_{0.10}$ with pure Ni. Since $x=0.11$ was made with $^{58}$Ni, this sample cannot be used for this comparison. We performed a polarized SANS measurement at a different instrument, VSANS, for a Ni sample that was prepared with the same annealing protocol as the alloys. We covered a large $Q$ range at room temperature, 300 K, where Ni is in the ferromagnetic phase ($T_c$=630 K). A stronger magnetic field of 0.5 T is needed for the Ni samples to avoid neutron beam depolarization. No clear net magnetic intensity results from the spin-flip contrast (SFH-SFV) with sufficient resolution that compares to the data of $x=0.10$. The upper limit is well below the cluster contribution of the alloys. However, we resolve a significant magnetic response from the DIF term, the flipper contrast. To better compare the observed intensities of the alloy Ni$_{0.90}$V$_{0.10}$ and of Ni, measured under different conditions, we calibrate the signal with the nuclear scattering term at low $Q$. In both samples we expect the nuclear scattering to be dominated by the same (natural) Ni material with the same Ni coherent scattering. So we calibrate the observed intensity DIFV=$2NM_x$ by dividing through the observed nuclear response $N^2$. This signal is measured by NSFH (after BG correction using the empty sample holder). We use here the value of $N^2$ at $Q_{\textrm{low}}=0.1$ nm$^{-1}$ that is part of the $1/Q^4$ Porod term as calibration constant $n_c=N^2(Q_{\textrm{low}})$.  
The calibrated $2NM_x(Q)/n_c$  should then be independent of the spectrometer configuration and sample size. 

\begin{figure}
\includegraphics[width=1\linewidth]{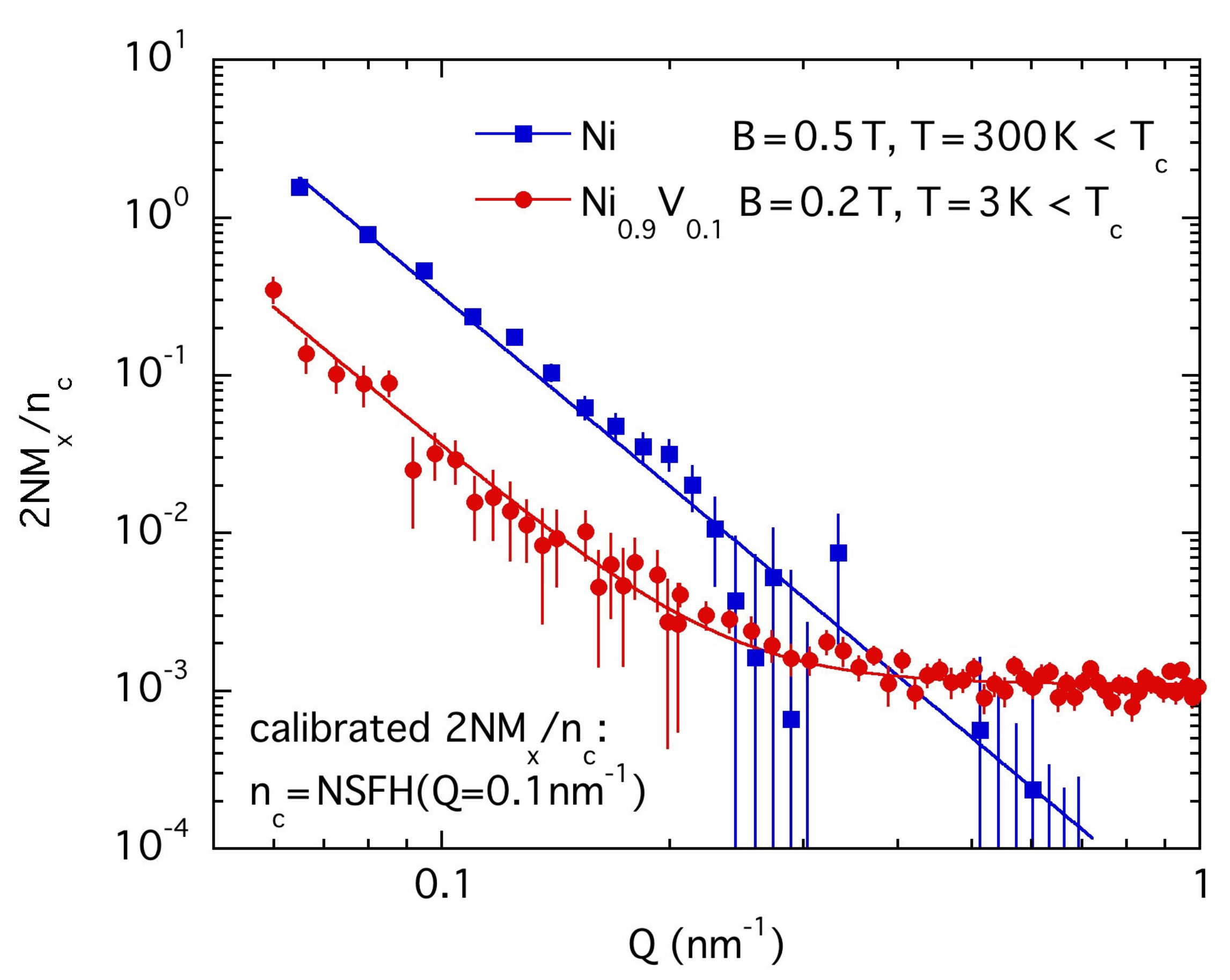}% nDIF0v10
\caption{ Calibrated DIF response $2NM_x/n_c$  for  Ni and Ni$_{0.90}$V$_{0.10}$ vs. wave vector $Q$. In Ni only the long-range ordered domain term is present, the local defects appear in the V alloy. The solid line is fit according to Eq.\ (\ref{eq:DIFV}).
}
\label{fig:DIFx0x10}
\end{figure}

Fig.\ \ref{fig:DIFx0x10} compares the $Q$ dependence of the calibrated DIF term, $2NM_x/n_c$, of  Ni$_{0.90}$V$_{0.10}$ and Ni  at the lowest possible fields. 
We see clearly for both samples at low $Q$ a dominating term $A_D/Q^4$, the amplitude is higher in pure Ni than in Ni$_{0.90}V_{0.10}$. At high $Q$, $2NM_x/n_c$ in Ni dips below the signal in the alloy, the extra constant $C_D$ term of the alloy (in Eq.\ (\ref{eq:DIFV})) seems missing in Ni. 
That supports that these local magnetic defects are caused by vanadium in Ni$_{0.90}$V$_{0.10}$ and not already present in Ni due to other structural defects or domain wall boundaries.  

\begin{figure}
\includegraphics[width=0.9\linewidth]{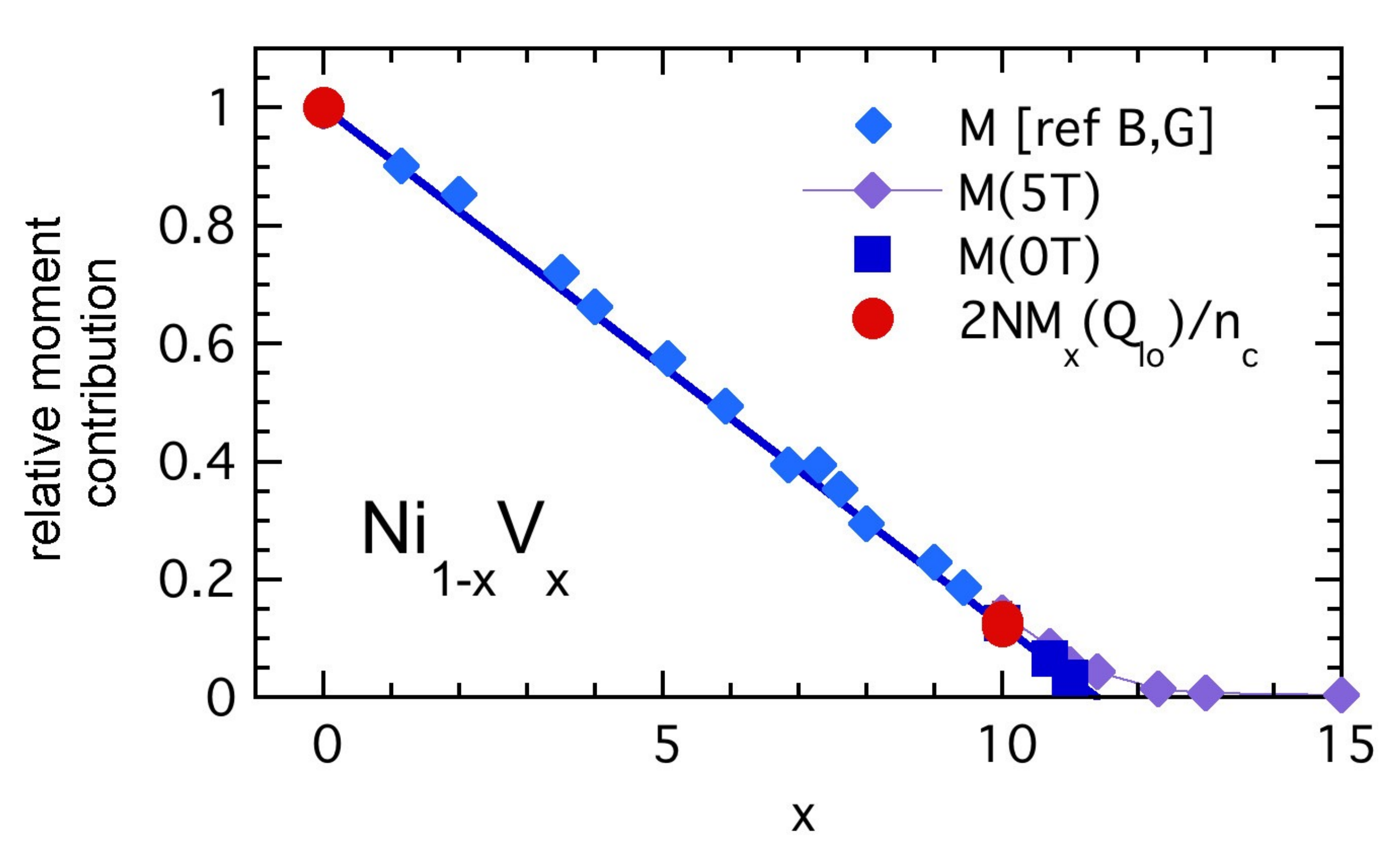}% xdep
\caption{ Relative moment contribution vs. vanadium  concentration $x$ from SANS results and other methods from Ref. \onlinecite{Wang2017}. $2NM_x(Q_{\textrm{low}})/n_c$ and bulk magnetization data $M(B)$ normalized to the value of Ni are shown. The long-range ordered domain term follows the trend of the linear reduction of the average magnetic moment with $x$.}
\label{momentdefect}
\end{figure}

Fig. \ref{momentdefect} summarizes the magnetic cross term $2NM_x/n_c$ at $Q_{\textrm{low}}$ at different V-concentration $x$. 
For better comparison ratios are shown normalized to $x=0$.
We see that the two SANS data points follow well  the bulk magnetization data from previous study \cite{Ubaid2010}. $M$(5T) represents the magnetization at high fields and $M$(0T) the spontaneous magnetization, the extrapolated value for 0T. The slight magnetic field dependence does not affect much the ratio. (The ratio of the bulk moment estimate of $x=0$ and $x=0.1$  increases from 0.13 in $B=0$ T to 0.14 in $B=5$ T, while the $2NM_x/n_c$ ratio estimate is $0.12$ in 0.2T and $0.13\pm0.02$ in 1.5 T). 
%Panel (b) illustrates how the constant high $Q$ DIF term  $2NM_x(Q_{hi})/n_c$ evolves with $x$ (with $Q_{hi}=0.6 $ nm$^{-1}$). 
 % There is no dominant magnetic $2NM_x/n(Q_{hi})$ term in pure Ni compared to the alloy, the ``$x=0/x=0.1$" ratio of $2NM_x(Q_{hi})/n_c$ is $0\pm0.3$.  That supports that Ni does not already contain magnetic local defects, they are caused by vanadium in Ni$_{0.90}$V$_{0.10}$. This rules out domain walls and other sources for such magnetic signal.

It is remarkable that the magnetic moment estimates from bulk data $M(B)$ and from SANS, extracting the Porod term $2NM_x(Q_{\textrm{low}})$, agree so well with each other. This DIF term does not only signal a magnetic response at low $Q$, it represents long-range ferromagnetic domain scattering. It even seems to measure the magnetic moment density and to function as an order parameter in this alloy. 
This cross term signal is so dominant since it is enhanced through the strong nuclear scattering of grain sizes in these polycrystalline samples. On one hand this is a huge ``background" obstructing the SF scattering at the low $Q$ range (that limits the precision of the magnetic short-range fluctuations), on the other hand this low $Q$ upturn does provide useful information.  
This simple scaling and successful calibration also imply that the magnetic structure and crystalline structure of Ni do not change very much in the alloy Ni-V, at least in these samples with low V concentrations prepared with the same protocol. Not only the cross sections but also the domain and crystallite structure determine the small angle scattering response.
The Porod term with amplitude $A_D$ typically depends on the contrast $(b_1-b_2)^2$ of the scattering length densities $b_1$ and $b_2$ but also on the surface properties and area $S$ in between the different scattering centers. Proper calibration with the nuclear term ($n_c=N^2$) assumes that the grain boundaries do not change much in the alloy compared to Ni. This agrees with a structural PDF analysis that notices minor defects in Ni but does not find changes in crystalline quality upon alloying \cite{Wang2017,PDF}. The fact that the $2NM_x(Q_{\textrm{low}})/n_c$ terms scale as the spontaneous magnetization $M$(0T) term suggests that also the FM domain sizes and boundaries between FM domains and potential PM inclusion do not differ much in the diluted alloys and in pure Ni.  Such structural and magnetic characterization is possible to extract from a simple cross term DIFV($Q$) that presents only one fit parameter: the amplitude of the Porod term $A_D$. It works because Ni and Ni$_{0.90}$V$_{0.10}$ contain the same nuclear scattering centers (natural Ni) and magnetic scattering centers. The present data from the Ni$_{0.89}$V$_{0.11}$ samples made with the $^{58}$Ni isotope cannot be easily used for this simple comparison. Further studies become necessary to study the domain scattering at higher vanadium concentration closer to $x_c$ when the fluctuations become more dominant and the remaining ordered network weakens. 

\subsection {\label{sec:x}Evolution of Magnetic Clusters with Disorder}
\begin{figure}
\includegraphics[width=0.9\columnwidth]{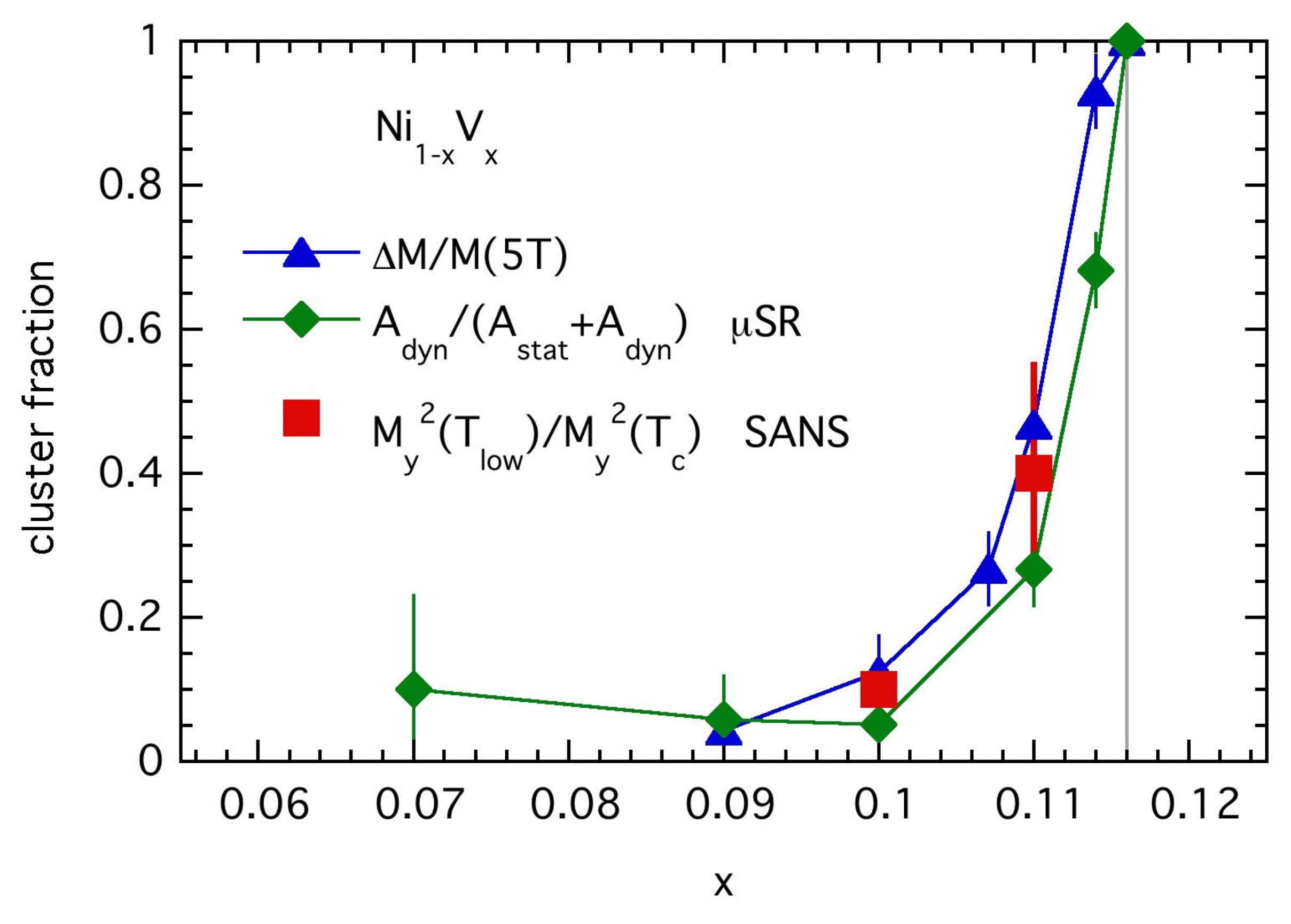}% ClusterRatio
\caption{Magnetic cluster fraction vs. vanadium concentration $x$ in Ni$_{1-x}$V$_x$ obtained from SANS compared to other methods: magnetization $M(B)$, $\mu$SR from Ref. \onlinecite{Wang2017}. The SANS estimate is the ratio of the SF contrast at lowest $T$ and $T_c$, taken in medium $Q$ range. 
}
\label{fig:ClusterRatio}
\end{figure}

The contribution of FM ordered moments measured by the domain term decreases with introducing vanadium into the alloy Ni$_{1-x}$V$_x$. At the same time some short-range magnetic fluctuations seem to develop in the FM state toward the critical V concentration $x_c$ before the long-range order breaks down. We measure the magnetic cluster contribution by the left over SF contrast at low $T$. To assess a cluster fraction we take the ratio of the signal intensity at lowest $T$ and at $T\approx T_c$ where all moments should be still in the PM phase.  
Fig.\ \ref{fig:ClusterRatio} shows the cluster fraction of the SANS measurement comparing with  previous estimates from  $\mu$SR  and magnetization \cite{Wang2017} as a function of $x$. 
The magnetization data define the power law increase in field up to 5 T as cluster contribution $\Delta M$. The $\mu$SR analysis recognizes an additional dynamic term as paramagnetic cluster contribution besides a static term describing the FM response. All these estimates agree well within their precision confirming the evolution of increasing magnetic fluctuations breaking the FM order at $x_c=0.116$. All data were taken at the lowest temperature of about 2 K. The present SANS data include data at 3 K and 4 K and non polarized data taken at 1.5 K. The cluster ratio for $x$=0.11 is slightly $T$-dependent, but not vanishing. The error bar includes possible uncertainties due to low $T$ extrapolations. 
A two component picture of PM and FM contribution is certainly oversimplified, but the present new SANS data rectify that extreme scales of long-range domains and a short-range scale of few nm are relevant in this alloy.  
Generally, it is clear that disorder renders the exchange interaction which ultimately leads to the emergence of the itinerant ferromagnetic state to become spatially inhomogeneous. As a result, one may expect that in a region of a sample, the local value of the exchange coupling would exceed the critical value of the averaged interaction constant leading to the formation of the ferromagnetic droplet. In two spatial dimensions, the size of the droplet does not depend on a size of the magnetic moment\cite{StonerDrops}, while in the three dimensional case this is not so. In either case, the dynamics of these droplets will be determined by the correlators of the randomly distributed coefficients in the Ginzburg-Landau functional. We leave the analysis of the magnetic susceptibility and heat capacity for future studies. 

%%%%%%%%%%%%%%%%%%%%%%%%%%%%%%%%%%%%%%%%%%%%%%%%%%%%%%%%%%%%%%%%%%%%%%%%%%%%%%%%%%%%%%%%%%%%%%%%%%%%%
% Conclusions
%%%%%%%%%%%%%%%%%%%%%%%%%%%%%%%%%%%%%%%%%%%%%%%%%%%%%%%%%%%%%%%%%%%%%%%%%%%%%%%%%%%%%%%%%%%%%%%%%%%%%
\section{Conclusion}
Although the average magnetic moment in Ni$_{1-x}$V$_{x}$ becomes very small close to the critical point where FM order vanishes, this SANS study identifies successfully magnetic scattering. 
The polarized SANS technique helps in particular to extract the small magnetic signal from the huge nuclear scattering. The pure spin flip (SF) response and the flipper difference (DIF) reveal complementary magnetic contributions at various length scales in Ni$_{1-x}$V$_{x}$ close to critical concentration $x_c$ where the onset of ferromagnetism at $T_c$ extrapolates to zero. 
The two samples with vanadium concentration of $x=0.10$ and $x=0.11$ present similar responses in the accessible $Q$-regime of (0.06-0.1) nm$^{-1}$ and are analyzed using simple forms. The result of a pure Ni sample serves as reference for a polycrystalline FM without atomic disorder.
 
First, we confirm the presence of magnetic clusters at the lowest temperatures far below $T_c$. The angular and $Q$ dependence (of the SF response) reveal isotropic magnetic short-range correlations similar to the paramagnetic fluctuations seen close to $T_c$. The scale of these correlations is 5-10 nm estimated by a simple Lorentzian function. %skip details
%The medium $Q$ response is strongest at the critical temperature $T_c$ as expected for critical fluctuations, but does not vanish completely towards low temperatures and therefore characterizes the remaining short-range magnetic fluctuations. 
These fluctuations get suppressed gradually in higher fields and follow qualitatively the expectations of the quantum Griffiths phase where magnetic clusters freeze with different fluctuations rates. The SF data give an estimate of the fraction of these magnetic clusters that do not contribute to the magnetic order. The cluster ratio is determined through the remaining few clusters at low $T$ in the FM phase compared to the PM signal close to $T_c$.
This cluster contribution increases significantly from $x=0.10$ to $x=0.11$, demonstrating that these short-range correlated fluctuations become more important toward $x_c$. The new SANS data agree well with previous cluster estimates from magnetization and $\mu$SR data \cite{Wang2017} supporting further the different dynamics of these contributions. 

In addition, we find signatures of long-range domain scattering, which supports macroscopic long-range order for samples even very close to $x_c$. Within this $Q$-range we cannot measure the macroscopic length scale, but only set lower limits and collect supporting evidence. The neutron beam gets depolarized below $T_c$ for both samples. The magnetic cross term DIF=$2NM_x$ shows a low $Q$ upturn due to long-range domain scattering with aligned moments $M_x$ similar to the pure nuclear scattering term $N^2$ due to grain boundary scattering.  This low $Q$ magnetic domain term only appears below $T_c$ and is field dependent. In higher magnetic fields, the low $Q$ domain term increases while the medium $Q$ fluctuation term decreases, indicating that the short-range fluctuations freeze to join the larger domains with aligned moments. A comparison with Ni even shows that this low $Q$ contribution scales with the magnetic moment estimates from bulk magnetization measurements. So, this low $Q$ domain term not only confirms long-range order but also reveals the ordered moment of the compound.

%The polarized SANS data reveal another magnetic anomaly that becomes most obvious in high magnetic fields. Already the SF data notice that the short-range response does not stay isotropic, a longitudinal component remains and does not decline like the transverse components in higher fields. The magnetic cross term demonstrates without doubt that there is another magnetic contribution visible at higher $Q$ that does not present any distinct $Q$ dependence up to 1nm$^{-1}$.  We confirm that such term is not present in a pure Ni compound. 
%The $Q$-independent magnetic contribution we found left over in higher fields is therefore caused by the vanadium that disturbs the magnetic homogeneity of the sample and is not related to any domain boundary effects or structural defects.
 
These SANS data provide remarkable information although the individual data lack precision in resolving details of a system with moments of less than $0.1\mu_B$/Ni. 
We see clear evidence of 
%less about local disorder: effective disorder and 
various magnetic correlations in this alloy produced by random atomic dilution. We notice 
%local magnetic defects produced by V and 
correlation lengths from nm scale to $\mu$m employing simple tools. 
%What are the consequences for the mechanism of a quantum critical point in a FM? How does this disordered FM get destroyed? 
These various contributions confirm that the FM order does not get homogeneously destroyed. 
%leading to a first order transition as expected for ideally clean non-disordered systems \cite{Brando2016}. 
On the other hand we do not notice any signs of a spin glass or cluster glass with only frozen short-range correlations. We do not see macroscopic phase separation either of different ordered phases. 
Note that a SANS response with a Porod or Lorentzian square term indicating longer-range domains and a shorter-range Lorentzian term is often observed in alloys and transition metals oxides (see e.g. in Ref \onlinecite{mSANSrev2019}). Typically, the size of the long-range order is already limited \cite{Hellman1999} or gets reduced by chemical substitution before the magnetic order breaks down  \cite{DeTeresa2006}. The two coexisting phases are expected both to be magnetic ordered phases with relevant coercivity \cite{ElKhatib2010}. 
We come to a different conclusion in our simple {\it d} metal alloy Ni$_{1-x}$V$_{x}$ with sustained long-range order and coexisting fluctuations close to $x_c$. Our prepared samples remain a soft metal and do not show any hysteresis in $M(B)$ \cite{Ubaid2010}.
The SANS data reveal macroscopic domains that are unavoidable in a finite sized soft FM.  We fit only two contributions of FM domains and PM clusters to facilitate a description of a variety of length and time scales present in these samples. It is likely that through quenched disorder the correlated regions vary throughout the alloy, leading to a network of longer-range domains that are perforated by shorter-range fluctuations (provoked by the random location of the V atoms). This disorder can lead to a quantum critical point in a FM metal. Here we tried to measure a typical scale of short-range correlations that stems from a distribution. We cannot see the local arrangement or a pattern formation. A complimentary method  with spatial resolution in nm would be of high interest to perform to reveal the location, distribution and shape of clusters and of coexisting long-range domains. 
This simple model system Ni$_{1-x}$V$_{x}$ with controlled random atomic distribution offers a great opportunity to study the complex interaction of a {\it d} metal with short and long-range interaction including local random defects. The present SANS data allow insight in the mechanism of  how fluctuations evolve in the FM ordered state that lead to a quantum critical point in an itinerant FM.

%%%%%%%%%%%%%%%%%%%%%%%%%%%%%%%%%%%%%%%%%%%%%%%%%%%%%%%%%%%%%%%%%%%%%%%%%%%%%%%%%
%problem with command ?????
%\ack
\section{Acknowledgement}
We thank J. Kryzwon, T. Dax, S. Watson and T. Hassan for their support with NG7SANS, cryogenics and $^3$He spin filters preparation at NIST. We thank Lisa DeBeer-Schmitt for her support at GPSANS, ORNL. 
This research is funded in part by a QuantEmX grant from ICAM and the Gordon and Betty Moore Foundation through Grant GBMF5305 to Hector D. Rosales. H.A. acknowledges support from Jazan University. M.D. acknowledges the financial support by the National Science Foundation grant NSF- DMR-2002795.
We acknowledge the support of the National Institute of Standards and Technology, U.S. Department of Commerce, in providing the neutron research facilities used in this work. Support for usage of the {$^3$He} spin polarizer on the NG7 SANS instrument was provided by the Center for High Resolution Neutron Scattering, a partnership between the National Institute of Standards and Technology and the National Science Foundation under Agreement No. DMR-1508249. A portion of this research used resources at the High Flux Isotope Reactor, which are DOE Office of Science User Facilities operated by Oak Ridge National Laboratory.
%\end{acknowledgments}

S.B. and H.A. contributed equally to this work.
%%%%%%%%%%%%%%%%%%%%%%%%%%%%%%%%%%%%%%%%%%%%%%%%%%%%%%%%%%%%%%%%%%%%%%%%%%%%
%

\appendix
\section{\label{sec:depol} Neutron depolarization}
The main study concentrates on the ``scattered" neutron intensities from the magnetic sample using the different cross sections with selected neutron polarization. The polarizing devices such as the supermirrors, spin flipper and $^3$He spin filter, where the polarizing efficiency decays with time, are certainly not perfect and do not produce a 100\% polarized beam. Collecting the transmitted direct beam intensity for all 4 spin combinations gives typically sufficient information for correction of such spin leakage. Another source for the depolarization of the neutron beam is the magnetic sample. On one hand depolarization measurements provide important information\cite{Halpern1941, Rekveldt1973, Mitsuda1985}, they can help to diagnose the magnetic state of the sample in particular for ferromagnets\cite{Yusuf2000, Mirebeau1991}. On the other hand a strong depolarization mixes and affects the different scattered intensities of the 4 cross sections that might make a proper polarization analysis unresolvable\cite{Quan2021}. The polarization state $P_S$ of the transmitted neutron beam due to the sample is essentially measured by the contrast of transmitted direct beam without spin flip $T_{\textrm{NSF}}$ and with spin flip $T_{\textrm{SF}}$ (for perfect polarizing devices with initial $P_0=1$).
\begin{equation}
\label{PfromT}
P_s=[T_{\textrm{NSF}} - T_{\textrm{SF}} ] / [T_{\textrm{NSF}} + T_{\textrm{SF}}]
\end{equation}
$P_s$ remains 1, when no change in the initial spin direction occurs. Conversely, $P_s$ goes down to 0, when the neutron spin looses completely its original orientation that the intensity of both initial and reversed direction are equal. 
\begin{figure}
\includegraphics[width=\columnwidth]{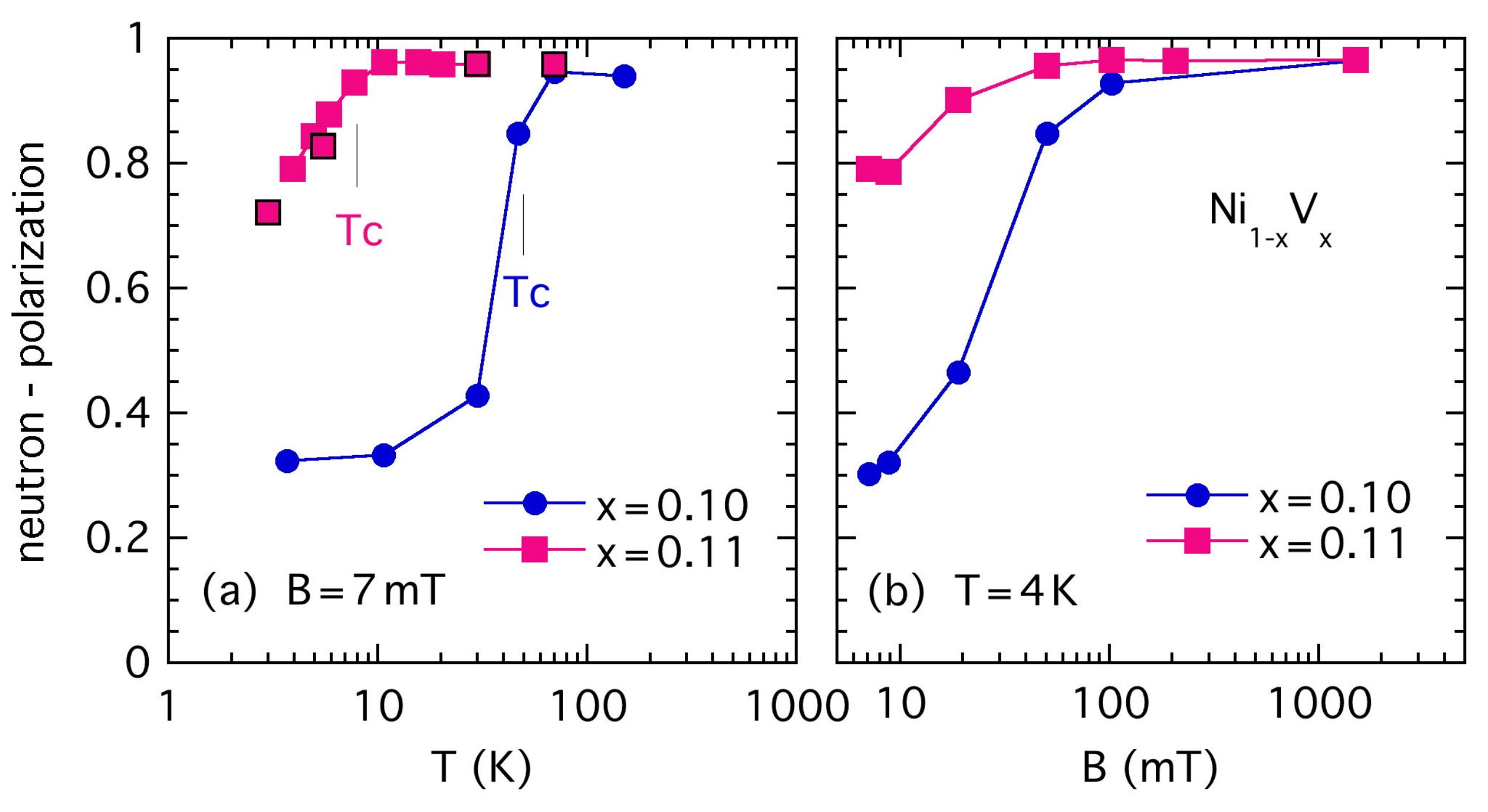}% depol  before Fig2 
\caption{Transmitted neutron beam polarization ratio $P_s$ of the sample (Ni$_{0.90}$V$_{0.10}$ and Ni$_{0.89}$V$_{0.11}$) at different environments. Panel (a) presents the temperature $T$ dependence in a small field, both samples depolarize the beam for $T<T_c$. Panel (b) shows how the polarization recovers in  applied magnetic fields $B$ at a low temperature.
}
\label{fig:depol}
\end{figure}

To check if the Ni-V samples depolarize the neutron beam and to find optimized working conditions to collect polarized scattering data we recorded the transmitted neutron beam polarization $P_s$ from the (efficiency corrected) transmission data at different temperature $T$ and magnetic fields $B$. Fig.\ \ref{fig:depol} presents $P_s$ for alloys with $x=0.10$ and $x=0.11$. $P_s$ is close to 1 for high temperatures in the paramagnetic region, but starts declining below the critical temperature $T_c$. Both samples are clearly depolarizing the beam in a small field of 7 mT. For $x=0.10$ $P_s$ saturates at low $T$ at a small value of $P_s=0.3$, for $x=0.11$ $P_s$ reaches 0.7 at 3 K. As expected the strong depolarization changes in higher fields. At a magnetic field of 1 T, the maximum value close to 1 is recovered, at 50 mT $P_s$ increases to 0.85 for $x=0.10$. We conclude that low $T$ scattering data taken at  50 mT show sufficient polarization for  Ni$_{1-x}$V$_x$ with $x\geq0.10$. 

Neutron depolarization is caused by strong internal fields that change on a mesoscopic scale. The different field directions make the neutron spin precess and deviate from the main polarization direction. Local defects or fast fluctuations in the sample do not affect the spin, but differently oriented long-range domains in a soft ferromagnet is the typical case that leads to depolarization as we are suspecting here. When these magnetic domains align in higher magnetic fields, the neutron polarization gets restored. The observed values of $P_s$ depend on the strength of the inner fields and the neutron time spend precessing to reach an effective precession angle $\phi$ and therefore on the dimensions, the domain size $\delta$ and the sample thickness $d$. We use a simple model proposed in Ref. \onlinecite{Yusuf2000} to estimate a typical domain size from $P_s$: 
\begin{equation}
\label{equation:depol}
P_s=P_f/P_i=\exp\left(-\frac{d}{3\delta}\phi_{\delta}^2\right),
\end{equation} 
where $P_f$ and $P_i$ are the final and initial neutron beam polarization. 
The precession angle $\phi_{\delta} =(4.63 \times10^{-10}G^{-1} \AA^{-2}) \lambda B \delta$ depends on neutron wave length $\lambda$, magnetic field $B$ and the domain size $\delta$. $B = 4 \pi M_s \rho$ can be estimated from spontaneous magnetization $M_s$ with density $\rho$ (all expressions in cgs units \cite{Yusuf2000}). For $x=0.10$ and $x=0.11$ we find $B=810$ G and $B=190$ G from the magnetization data \cite{Wang2017}, which yield domain sizes of $3 \mu$m and $2 \mu$m, respectively. 
This estimate of similar domain sizes is certainly too crude for the present alloys, but these depolarization data demonstrate that both samples show all indications of ferromagnetic order at low  $T<T_c$ with long-range domains on the order of $\mu$m. 
%

%\bibliographystyle{apsrev4-1}
%\bibliography{PSANSv4.bib}

%\end{document}

%

\end{document}